\documentclass[final,times,1p]{elsarticle} 
\usepackage{amssymb}
\usepackage{amsmath}
\usepackage{xurl}
\usepackage{tikz}
\usepackage{soul}
\usetikzlibrary{arrows.meta}
\usepackage{hyperref}
\usepackage{setspace} 

\usepackage{xcolor}

\journal{Physica A: Statistical Mechanics and its Applications}

\begin{document}

\begin{frontmatter}

\title{On the potential of quantum walks for modeling financial return distributions}

\author[inst1]{Stijn De Backer}
\ead{stijn.debacker@ugent.be}

\affiliation[inst1]{organization={Department of Physics and Astronomy},
            addressline={Proeftuinstraat 86}, 
            city={Ghent},
            postcode={9000}, 
            country={Belgium}}

\affiliation[inst2]{organization={Department of Economics},
            addressline={Sint-Pietersplein 5}, 
            city={Ghent},
            postcode={9000}, 
            country={Belgium}}

\affiliation[inst3]{organization={Department of Economics},
            addressline={Tweekerkenstraat 2}, 
            city={Ghent},
            postcode={9000}, 
            country={Belgium}}
            
\author[inst1,inst2]{Luis E. C. Rocha}
\author[inst1]{Jan Ryckebusch}
\author[inst3]{Koen Schoors}

\begin{abstract}
Accurate modeling of the temporal evolution of asset prices is crucial for understanding financial markets. We explore the potential of discrete-time quantum walks to model the evolution of asset prices. Return distributions obtained from a model based on the quantum walk algorithm are compared with those obtained from classical methodologies. We focus on specific limitations of the classical models, and illustrate that the quantum walk model possesses great flexibility in overcoming these. This includes the potential to generate asymmetric return distributions with complex market tendencies and higher probabilities for extreme events than in some of the classical models. Furthermore, the temporal evolution in the quantum walk possesses the potential to provide asset price dynamics. 
\end{abstract}


\end{frontmatter}

\section{Introduction}
\label{section:introduction}

The analysis of the price fluctuations in financial time series is of major interest in economics and finance. It contributes to our understanding of the functioning of financial markets, and how they can be regulated. Related research in econophysics concentrates on applying techniques from statistical physics, complexity science and network theory~\citep{Sharma2011}. Thereby, one predominantly adopts an approach based on the concepts of classical physics and rarely touches on the rich variety of methodologies that characterize quantum physics. The latter is often associated with the description of physical properties at the atomic and subatomic scale, but the underlying mathematics offer a framework for interpreting measurements~\citep{pitowsky2006quantum} and propagating uncertainty, which finds applications in a variety of fields~\citep{busemeyer2015quantum}. Moreover, quantum physics introduces concepts such as entanglement that are absent in classical physics but can enrich other fields~\citep{orrell2018quantumbooknewscience, Orrell:MoneyMagic}.  

A research field that has seen the exploration of quantum-inspired methodological extensions is human cognition theory. Upon examining the processes of judgment and decision-making, which are also fundamental processes in behavioral economics, there is a tendency to rely on principles taken from classical computational logic, classical dynamic systems theory, and classical probability theory~\citep{busemeyer2015quantum}. Quantum cognition theory is an interdisciplinary research field that leverages quantum theory methodologies as a framework for comprehending human cognition. It departs from classical assumptions, for example by allowing interference effects~\citep{pothos2022quantum}. Multiple experiments have provided evidence for the anticipated interference effects~\citep{hampton1988disjunction, shafffi1990typicality, tentori2004conjunction, aerts2012quantum}, highlighting the significance of exploring beyond the conventional classical frameworks in cognition theory. 
 
The potential of quantum-inspired methodologies in an economic context is exemplified in some recent works. For example, Orrell designed a model for supply and demand which features a classical approach and a quantum alternative based on the quantum harmonic oscillator~\citep{orrell2020supplyanddemand, orrell2022quantum}. In the quantum approach, supply and demand are represented as probabilistic wave functions in accordance with their uncertain nature, which contrasts with the deterministic perspective in a classical setting. In another example, Ahn et al.~modeled the temporal evolution of financial return distributions with the Fokker-Planck equation and derived the Schrödinger equation for a particular external potential, eventually suggesting power law behavior in the tail of the return distributions~\citep{ahn2024business}. Thereby, the value of the power law exponent can be associated with business cycles. Meng et al.~modeled the stock market using a quantum version of Brownian motion in order to study non-Markovian behavior of stock indices~\citep{meng2016quantum}. 

An advantage of using quantum-inspired algorithms in an economic setting is the fact that they may offer practical advantages in their algorithmic implementation. Indeed, quantum computers hold the potential of significantly revolutionizing the handling of computationally challenging problems by outperforming classical computers for selected tasks~\citep{orus2019quantum, egger2020quantum}. This makes quantum computing an attractive area to solve computational problems in finance and economics~\citep{herman2022survey, widdows2023quantum, widdows2024quantum}. It is important to note that a quantum circuit for a one-dimensional quantum walk can be implemented efficiently on a quantum computing device~\citep{puengtambol2021implementation}.  

The inclusion of quantum based methodologies in economics and finance does not imply that economic systems and financial markets should be understood and interpreted as being quantum systems. Sometimes, it can be tempting to envision a singular, universal model to capture the price evolution of financial assets and the probability distributions of stock returns. Econophysics offers a variety of models, each having its strengths and weaknesses in terms of capturing the asset price evolution and the observed market behaviors. The application of methodologies originating from quantum physics adds to this diversity. 

In this paper, we explore the potential of discrete-time quantum walks in modeling the temporal evolution of the price of financial products and the resulting return distributions. The discrete-time quantum walk can be looked upon as the quantum counterpart of the classical random walk~\citep{venegas2012quantum}.  The goal of this paper is to provide a comparative study between a model for the price evolution of financial products based on the discrete-time quantum walk, and state-of-the-art classical methodologies, including geometric Brownian motion~\citep{BlackScholes:1973, ziemann2021physics}, Lévy stable distributions~\citep{mandelbrot1963variationcertain}, and power law distributions~\citep{gabaix2009power, gopikrishnan1998inverse, gopikrishnan1999scaling}. We investigate the pros and cons of a quantum approach in terms of capturing the diffusion properties of financial assets and the distribution of the fluctuations about general trends. The findings may offer valuable insights into the assessment of financial risk.

The addition of interference effects and quantum entanglement to model asset price evolution through the concept of quantum walks comes at the expense of the introduction of a number of parameters. Throughout this paper, we pay attention to the connection of those parameters to specific market dynamics and provide a systematic study of the parameter space to show their impact on the resulting distributions. An important aspect is to explore whether the advantages of the quantum walk model in capturing the characteristics of financial return distributions justify the added complexity and suggest promising performance from a theoretical perspective. This paper focuses on the theoretical potential of quantum walks to model the temporal evolution of asset prices. A detailed analysis of the model's performance against real-world asset price returns will be explored in future work.

The structure of this paper is as follows. Section \ref{Section:Models} overviews the basic principles of the classical models for the financial return distributions, followed by an examination of the quantum walk algorithm and its application to simulating the statistical properties of financial time series. In Section \ref{Section:Results}, we provide a financial interpretation of the parameters governing the quantum walk dynamics, and highlight how the quantum approach can overcome specific issues encountered in the classical methodologies. Finally, Section \ref{Sec:Conclusions} summarizes our main conclusions.

\section{Models for the temporal evolution of financial asset prices}   \label{Section:Models}

\subsection{Geometric Brownian motion}
One of the simplest models for the price evolution of financial assets is based on a classical random walk or geometric Brownian motion (GBM). This model assumes that price fluctuations above the general trend are random, resulting in a mathematical expectation of zero for the profit~\citep{Bachelier:1900, orrell2020quantummathsbook}. Stock prices change under the influence of random impacts, which are the financial equivalent of collisions with physical particles in Brownian motion.

The share price increments $dS = S(t + dt) - S(t)$ can be modeled using GBM~\citep{ziemann2021physics} 
\begin{equation} \label{GeometricBrownianMotion}
    dS = \mu S(t) dt + \sigma S(t) dW(t) ~,
\end{equation}
where $S(t)$ is the share price at time $t$, $dt$ is an infinitesimal time interval, $\mu$ is the average drift rate, $\sigma$ is the volatility, and $W(t)$ represents a white-noise Wiener process. The successive increments $dW(t)$ in the stochastic term are random, statistically independent, identically distributed and have a Gaussian nature~\citep{mandelbrot1997variation, plerou2000econophysics}. The first term on the right-hand side of Eq.~(\ref{GeometricBrownianMotion}) corresponds with a drift term. Setting $\mu=0$ corresponds to a situation with no general drift, leaving only the fluctuations. Using Itô's lemma, the share price at time $t$ can be related to the initial price $S(0)$ by~\citep{mao2007stochastic}
\begin{equation}
    S(t) = S(0) \exp{\left[ \sigma W(t) + (\mu - \sigma^2/2)t \right]} ~.
\end{equation}

We define the normalized logarithmic return over a time period $\Delta t$ as
\begin{equation}   \label{EqDefinitionNormalizedReturn}
    g = \frac{\log_e{S(t + \Delta t)} - \log_e{S(t)}}{\sigma} ~,
\end{equation}
where $S(t)$ is the price of the stock at time $t$, and its volatility $\sigma$ is the standard deviation of the distribution of $\left( \log_e{S(t + \Delta t)} - \log_e{S(t)} \right)$. The return distribution $P(g)$ is the probability distribution of the returns, which can be modeled using different approaches (e.g.~GBM). The GBM model of Eq.~(\ref{GeometricBrownianMotion}) assumes a constant value of the volatility parameter $\sigma$. The width of the return distribution grows with $t^{1/2}$~\citep{ziemann2021physics}. This feature will be compared to the diffusion properties of the discrete-time quantum walk. 

Although numerous shortcomings have been identified when comparing the predictions of the GBM model to financial data, the model holds a strong position in quantitative finance, and it is one of the underlying principles of the Black-Scholes model~\citep{BlackScholes:1973}. In Section \ref{Section:QuantumWalk}, we explore the potential of substituting the Wiener process in Eq.~(\ref{GeometricBrownianMotion}) with a discrete-time quantum walk in order to address some of the limitations of the GBM model. Some specific limitations include:

(i) The inability of the GBM model to capture large price changes which occur much more frequently than the Gaussian distribution predicts~\citep{yarahmadi20222d}. When examining return distributions on small time scales, one finds that the distribution is more peaked and the tails are heavier than the Gaussian assumption would permit~\citep{ding1993long}. This can be described as a ``leptokurtic'' feature of the return distributions. Bachelier acknowledged the existence of large price variations back in 1900, but categorized them as outliers without paying further attention~\citep{Bachelier:1900, mandelbrot1997variation}; 

(ii) GBM results in symmetric return distributions. For real data, however, the tails rarely display full symmetry~\citep{kou2002jump};

(iii) GBM is consistent with the efficient market hypothesis (EMH)~\citep{malkiel2003efficient, Holt:2013, Scanlon:2019}. In a competitive and perfectly informed market, the current speculative prices are typically assumed to permanently incorporate expected or predictable changes. Unpredictable and unforeseeable alterations, however, remain subject to speculation, and these changes are assumed to behave randomly. Consequently, a random process is considered the most suitable choice for modeling these fluctuations under the EMH. The latter has been vehemently criticized~\citep{malkiel2003efficient}, claiming that it should be, in fact, possible to ``beat the market''. Moreover, the concept of fluctuations in a financial market being entirely driven by a random process remains a topic of skepticism, especially the fact that the stochastic term in Eq.~(\ref{GeometricBrownianMotion}) does not account for temporal correlations~\citep{lekovic2018evidence}. Moreover, inferring knowledge from data becomes notably difficult under the assumption of independent increments.

\subsection{Lévy stable and power law distributions}
Mandelbrot argued that a Lévy stable distribution possesses a greater potential to model price increments~\citep{mandelbrot1963variationcertain, fama1963mandelbrot}. The Lévy stable distribution does not possess a closed-form expression for its probability density function (PDF) and cumulative distribution function (CDF). Its characteristic function is given by~\citep{mittnik1999maximum}
\begin{equation}    \label{Eq:Levystable}
    \varphi(t; \alpha, \beta, c, \mu) = \exp{\left( i \mu t - | ct | ^{\alpha} \left( 1 - i \beta \frac{t}{|t|} \omega(|t|,\alpha) \right) \right) } ~,
\end{equation}
where $0 < \alpha \leq 2 $ is the characteristic exponent or tail index, $-1 \leq \beta \leq 1 $ is the skewness parameter, $c>0$ is the scale parameter, $\mu \in \mathbb{R}$ is a shift parameter, and the function $\omega(|t|,\alpha)$ is defined by
\begin{equation}
    \omega(|t|,\alpha) = \left\{
    \begin{array}{ll}
        \tan{\frac{\pi \alpha}{2}} & \mbox{for} \ \alpha \neq 1 \\
        - \frac{2}{\pi} \log_e{|t|} & \mbox{for} \ \alpha = 1 ~.
    \end{array}
\right.
\end{equation}
The corresponding probability distribution function $f(x; \alpha, \beta, c, \mu)$ can be computed through
\begin{equation}
    f(x; \alpha, \beta, c, \mu) = \frac{1}{2 \pi} \int_{-\infty}^{+\infty} \exp(-ixt) \varphi(t; \alpha, \beta, c, \mu) dt ~.
\end{equation}
The parameter $\beta$ is associated with the asymmetry of $f(x; \alpha, \beta, c, \mu)$. For $\beta = 0$, the distribution is symmetric. For $\alpha = 2$ and $\beta=0$, one retrieves the Gaussian distribution $\varphi(t; 2, 0, c, \mu) = \exp{\left( i \mu t - (ct)^2 \right)}$ with mean $\mu$ and variance $2c^2$. Therefore, the model based on a Lévy stable distribution can be looked upon as an extension to the GBM model.

The Lévy stable distribution does not possess a finite value of the variance for $\alpha < 2$. This can pose a substantial challenge in a financial context, for example when analyzing return distributions. Mandelbrot argued that an infinite variance is appropriate to deal with the occasional large price changes~\citep{mandelbrot1966persistence}, a point of view that is heavily contested in the current econophysics literature~\citep{Sharma2011}.

CDFs of the power law type $\propto x^{- \tilde{\alpha}}$ have been suggested as an alternative way to deal with the occurrence of more extreme events, given their fat-tails like some Lévy stable distributions. Extensive financial data sets have been analyzed to investigate the CDFs of the returns defined over different time scales $\Delta t$. Gopikrishnan et al.~demonstrated by analyzing the S\&P 500 Index (USA), the Nikkei Index (Japan), and the Hang Seng Index (Hong Kong) that for \mbox{$\Delta t \lesssim 4$ days}, the CDF of the returns is consistent with asymptotic power law  behavior, where the exponent is given by $\tilde{\alpha} \approx 3$~\citep{gopikrishnan1999scaling}, which is beyond reach for the $0 < \alpha \leq 2 $ of the Lévy stable distributions~\citep{plerou2000econophysics}. This behavior is often referred to as the inverse cubic law of stock price fluctuations~\citep{gabaix2009power, gopikrishnan1998inverse}, which is applicable to returns defined over time periods varying from 15 seconds to a few days. For longer time scales, a slow convergence to Gaussian behavior is observed~\citep{meng2016quantum, madan1990variance, kiyono2006power, tuncay2007power, kiyono2006criticality}. 

In another analysis, Gopikrishnan et al.~investigated trades for all stocks in three major US stock markets in a period of two years~\citep{gopikrishnan1998inverse}. A major conclusion was that the asymptotic behavior of the CDF of the relative price changes defined over $\Delta t = 5$ min can be well-described by a power law with $\tilde{\alpha} \approx 3$. 

The conclusions drawn from studies yielding asymptotic power law behavior in the return distributions cannot be generalized to all financial markets. For instance, a similar data analysis on normalized price fluctuations ($\Delta t = 1$ day) in the National Stock Exchange in India over an 8-year period did not result in power law behavior~\citep{matia2004scale}, but in an exponential $P(g) \propto \exp{(-\beta g)}$. Apparently, the fluctuations in the Indian stock market are scale-dependent, which is not the case for markets exhibiting power law behavior. Another data analysis investigated the cumulative distribution of the normalized 1-min returns in the Shanghai Stock Exchange~\citep{zhang2007power}. It was concluded that due to irregular near-closure returns, neither the positive tail nor the negative tail can be fitted by power laws.

The above discussions suggest that there is no one-size-fits-all approach in the modeling of price fluctuations in return distributions.

Some limitations of the models for probability distributions of the stock returns based on Lévy stable and power law distributions include: 

(i) Convergence of the higher moments of Lévy stable and power law distributions is not guaranteed. The $m$th moment of a power law distribution with exponent $k$ exists for ${m < k-1}$, while all higher moments diverge~\citep{newman2005power}. This is rather impractical upon analyzing real-life financial data.

(ii) Models with an imposed functional form of the return distribution do not offer an in-depth understanding of the generative mechanisms governing the observed temporal evolution of asset prices. The generative mechanisms for power law behavior are an active area of research~\citep{ahn2024business, gabaix2009power, gabaix2003theory, farmer2004origin, mitzenmacher2004brief, markovic2014power}. Upon modeling the temporal evolution of asset prices with a discrete-time quantum walk, one can gain a comprehensive view of the internal dynamics. Indeed, one has full control over the parameters that govern the dynamics of the quantum walk and the resulting probability distributions. This approach allows one to tune the model parameters so that the resulting statistical distributions match those of the observed financial assets. In Section \ref{Section:QuantumWalk}, we provide a detailed discussion of these features.

Before moving to the discrete-time quantum walk, we mention that there exists a plethora of classical methodologies for modeling financial time series. Many adaptations of the basic GBM model (Eq.~(\ref{GeometricBrownianMotion})) have been proposed, for example the ``constant elasticity of the variance'' (CEV) model~\citep{yuan2018cev}, or a model using fractional Brownian motion~\citep{rogers1997arbitrage, rostek2013note}. Other noteworthy classes of models to address return distributions include jump diffusion processes~\citep{kou2002jump, merton1976option, hanson2002jump}, exponentially truncated Lévy distributions~\citep{cont1997scaling}, hyperbolic distributions~\citep{eberlein1995hyperbolic}, normal inverse Gaussian distributions~\citep{barndorff1997normal}, Student's $t$ distributions~\citep{blattberg2010comparison}, stretched exponential distributions~\citep{laherrere1998stretched}, and log-Weibull distributions~\citep{malevergne2005empirical}. 

\subsection{The discrete-time quantum walk} \label{Section:QuantumWalk}
In this paragraph, we introduce the concept of quantum walks to study the fluctuations in the temporal evolution of financial asset prices. Quantum walks have been used in a model for financial option pricing~\citep{Orrell:MoneyMagic, orrell2020quantummathsbook, orrell2021quantumwalkArticle}. Thereby, the probability distributions from a Hadamard quantum walk were introduced in an attempt to better reflect the economic agents' projections on the future value of the underlying asset. Economic decisions made by investors are seen as a collapse of their ``mental wave function'', which corresponds with a measurement process. In the case of financial options, however, investors' beliefs about the future may be more extreme than the reality, since such a collapse does not occur when taking a position about the future value of the underlying asset. The model was not generalized to the actual temporal evolution of the underlying securities of the option contract. Moreover, the model's reliance on a Hadamard coin and a symmetric initial configuration can be extended, which is one of the major goals of this paper. Before proceeding to the application of the quantum walk in a financial context and its economic interpretation, we briefly revise the fundamental algorithm underlying the quantum walk while focusing on the parameters that define its dynamics~\citep{venegas2012quantum, kempe2003quantum, chandrashekar2008optimizing, reitzner2012quantum}. 

\subsubsection{Formalism}
The discrete-time quantum walk algorithm was proposed in Ref.~\citep{aharonov1993quantum}. The state of the system lies in the Hilbert space $\mathcal{H} = \mathcal{H}_C \otimes \mathcal{H}_P$, where the two-dimensional coin space $\mathcal{H}_C$ is spanned by the two basis spin-1/2 states 
\begin{equation}
    \left| \uparrow \right\rangle = \begin{pmatrix} 1 \\ 
    0
\end{pmatrix},~ \left| \downarrow \right\rangle = \begin{pmatrix} 0 \\ 
1
\end{pmatrix}~.
\end{equation}
The position space $\mathcal{H}_P$ is spanned by a discrete set of the position basis states $\{ \vert j \in \mathbb{Z} \rangle \}$. One time step in the quantum walk algorithm consists of two separate operations. First, a quantum coin toss operator $\widehat{C}$ is applied on the coin part of the current state. The coin toss for a two-dimensional coin can be written as an SU(2) operator of the form
\begin{equation} \label{SU2coinDefinition}
    U_{\xi,\theta,\zeta} = \begin{pmatrix}
  e^{i\xi} \cos{\theta} & e^{i\zeta} \sin{\theta} \\ 
  e^{-i\zeta} \sin{\theta} & -e^{-i\xi} \cos{\theta}
\end{pmatrix}~,
\end{equation}
where $\xi$, $\theta$ and $\zeta$ are the Cayley-Klein parameters~\citep{chandrashekar2008optimizing}. The only relevant parameters for the dynamics of a quantum walker initialized at position $j=0$, are $\eta \equiv \xi - \zeta \in [0, 2 \pi[$ (proof in~\ref{sec:appendix}) and $\theta \in [0, \pi[$, since multiplying the $U_{\xi,\theta,\zeta}$ with a global phase does not impact the final result~\citep{jayakody2023revisiting}. In many practical applications, one sets $\xi = \zeta = 0$, leading to the following coin
\begin{equation} \label{UthetacoinDefinition}
      U_{\theta} \equiv  U_{\xi=0,\theta,\zeta=0} = \begin{pmatrix}
  \cos{\theta} &  \sin{\theta} \\ 
   \sin{\theta} & - \cos{\theta}
\end{pmatrix}~.
\end{equation} 
The Hadamard coin corresponds with $U_{\theta= \pi/4}$. After the operation of $\widehat{C}$, a conditional translation operator $\widehat{T}$ is applied, changing the position state to the left or to the right, depending on the result of the quantum coin toss. The unitary operator $\widehat{T}$ is defined by
\begin{equation} \label{DefinitionTranslation}
    \widehat{T} = \left[ \left| \uparrow \right\rangle \left\langle \uparrow \right| \otimes \left( \sum_{j \in \mathbb{Z}} \vert j + 1 \rangle \langle j \vert \right) \right] + \left[ \left| \downarrow \right\rangle \left\langle \downarrow \right| \otimes \left( \sum_{j \in \mathbb{Z}}  \vert j - 1 \rangle \langle j \vert \right) \right]~,
\end{equation}
and couples the temporal coin status to the positional status of the system. 

We define the unitary operator $\widehat{V}$ as
\begin{equation}    \label{EqOperotarV}
   \widehat{V} = \widehat{T} \cdot (\widehat{C} \otimes \widehat{\mathbb{I}}_{P})~,
\end{equation}
where $\widehat{\mathbb{I}}_{P}$ is the unity operator in position space. A quantum walk of $n$ discrete time steps consists of applying the operator $\widehat{V}^n$ to an initial state $\vert \psi (n=0) \rangle$. \ref{sec:appendix} provides additional details that offer deeper insight into the algorithm.

After $n$ time steps, the system's wave function can be written as 
\begin{equation}   \label{EqWaveFunction}
    \vert \psi (n) \rangle = \sum_{j = - n}^{ + n} \Bigl( a_j(n) \left| \uparrow \right\rangle + b_j(n) \left| \downarrow \right\rangle \Bigr) \otimes \vert j \rangle ~,
\end{equation}
where the coefficients $a_j(n)$ and $b_j(n)$ are introduced for the ``up'' and ``down'' components respectively. Throughout this paper, we systematically use $n$ to denote discrete time, and $j$ to denote spatial positions. 

The probability to measure the walker at position $j$ after $n$ time steps is given by the squared amplitude of the wave function (including both up and down components)
\begin{equation}    
\label{EqprobabilityQuantumWalk}
    P_j(n) = | \langle j \vert \psi(n) \rangle |^2 = | a_{j}(n) | ^2 + | b_{j}(n) | ^2 ~.
\end{equation}
The quantum walk distinguishes itself from the classical random walk by the fact that the randomness does not arise from a stochastic transition between states, but from the inherent unpredictability of the outcome of a quantum measurement process. Before measurement, the probability amplitudes $a_j(n)$ and $b_j(n)$ of all possible outcomes evolve in coincidence. The time evolution in the quantum walk is characterized by a superposition of all possible trajectories, allowing for interference in successive time steps and for quantum correlations. 

\subsubsection{Quantum walk model for financial time series}
We now introduce the model for the temporal evolution of the price of financial products. The central equation of the quantum walk model for financial time series is derived by replacing the Wiener process $W(t)$ in the stochastic differential equation of Eq.~(\ref{GeometricBrownianMotion}) by a quantum walk process $Q(t)$
\begin{equation} \label{EquationQuantumDiffusionForStocks}
    dS = \mu S(t) dt + \sigma S(t) \bigl( f(t) dQ(t) \bigr)~.
\end{equation}

An interesting feature of the quantum walk is that the variance of the resulting position probability distribution $P_j(n)$ (Eq.~(\ref{EqprobabilityQuantumWalk})) grows quadratically with the number of time steps $n$. This diffusive behavior is often referred to as ``ballistic diffusion''~\citep{romanelli2007measurements, ishak2021entropy}. In a classical random walk, the variance grows linearly with the elapsed time. For now, we have added an additional function $f(t)$ in the stochastic term to control the diffusion properties of the asset price evolution. This function is intended to account for the ballistic diffusion of the quantum walk, which one might seek to compensate for by setting $f(t) \propto t^{-1/2}$, resulting in more ``classical'' diffusive behavior. Indeed, the width of the quantum walk probability distribution increases linearly with time $t$, in contrast to GBM, where it scales with $t^{1/2}$. Note that this function does not have a quantum mechanical interpretation.

The function $f(t)$ can be chosen such that the standard deviation of the stock returns matches a given value of the stock's volatility $\sigma$. It is observed that the absolute values of the returns have a positive autocorrelation for time scales ranging from a few minutes to several weeks, whereas the actual values of the returns have non-vanishing but rather insignificant autocorrelations~\citep{ding1993long, farmer2000physicists, cont2007volatility}. Based on these observations of so-called volatility clustering~\citep{mandelbrot1963variationcertain}, the value of $\sigma$ is often kept fixed.

Volatility clustering provides an argument for the viability of the quantum walk model. The temporal evolution in the discrete-time quantum walk can be decomposed into Markovian and interference terms~\citep{romanelli2004quantum}. Neglecting the interference terms leads to a purely Markovian process. Volatility clustering suggests that asset returns do not follow an independent and identically distributed (i.i.d.) process, nor do any functions of the returns, such as their absolute values. In Refs.~\citep{ding1993long, malevergne2005empirical}, it is argued that the returns of the S\&P index are not governed by an i.i.d.~process. One can argue that the use of a quantum walk-inspired model better aligns with the non-Markovian properties observed in the temporal evolution of asset prices. In the forthcoming section, we provide an economic interpretation of the quantum walk in the context of financial time series, focusing on the parameters that shape the return distributions.

Table \ref{TableModelParameters} lists all parameters of the models discussed in this section. The interpretation of the quantum walk parameters will be the main topic of the forthcoming section. The role and the impact of the parameter $p$ will be discussed in Section~\ref{Sec:Decoherence}.

\begin{table}
\begin{tabular}{ |p{3cm}||p{2cm}|p{7cm}| }
\hline
\textbf{Model} & \textbf{Parameter} & \textbf{Interpretation} \\
\hline
GBM & $\mu$ & general trend, drift \\
(Eq.~(\ref{GeometricBrownianMotion})) & $\sigma$ & volatility, magnitude of the fluctuations \\
 & $W(t)$ & Wiener process, nature of the fluctuations \\
\hline
Lévy stable & $\mu$ & general trend, drift\\
distribution (Eq.~(\ref{Eq:Levystable})) & $c$ & scale parameter, magnitude of the fluctuations\\
& $\alpha$  &   tail index, asymptotic behavior, occurrence of extreme events \\
 & $\beta$ & skewness parameter, asymmetry, bias \\
\hline
Quantum walk & $\mu$ & general trend, drift \\
(Eqs.~(\ref{EqWaveFunction}), (\ref{EqprobabilityQuantumWalk}), (\ref{EquationQuantumDiffusionForStocks})) & $\sigma$ & volatility, magnitude of the fluctuations \\
 & $f(t)$ & function designed to control diffusion properties \\
 & $n$  &  number of time steps \\
 & $\vert \psi(0) \rangle$ & skewness, asymmetry, bias \\
 & $\eta$, $\theta$ & skewness, asymmetry, bias, diffusion properties \\
 & $p$ & level of transition to classical behavior \\
\hline
\end{tabular}
\caption{\label{TableModelParameters}Summary table with the discussed models and the interpretation of their parameters in financial terms.}
\end{table}

\section{Results} \label{Section:Results}
\subsection{The unitary quantum walk parameters in financial time series}
The quantum walk dynamics and the resulting probability distribution $P_j(n)$ of Eq.~(\ref{EqprobabilityQuantumWalk}) are controlled by the number of time steps $n$, the initial condition $\vert \psi (0) \rangle $, and the values of ($\eta = \xi - \zeta$, $\theta$) in the coin operator of Eq.~(\ref{SU2coinDefinition}). In this paragraph, we provide a mapping of these parameters on financial properties, considering the quantum walk as a modeling tool for financial time series (Eq.~(\ref{EquationQuantumDiffusionForStocks})), and we explore how they can be controlled to capture the behavior and characteristics of financial data, e.g.~market trends.

\subsubsection{Dependence on the number of time steps}
\begin{figure}[hbt!]
\begin{center}
\includegraphics[width=0.5\textwidth,height=2cm,keepaspectratio,angle=0,scale=5.2]{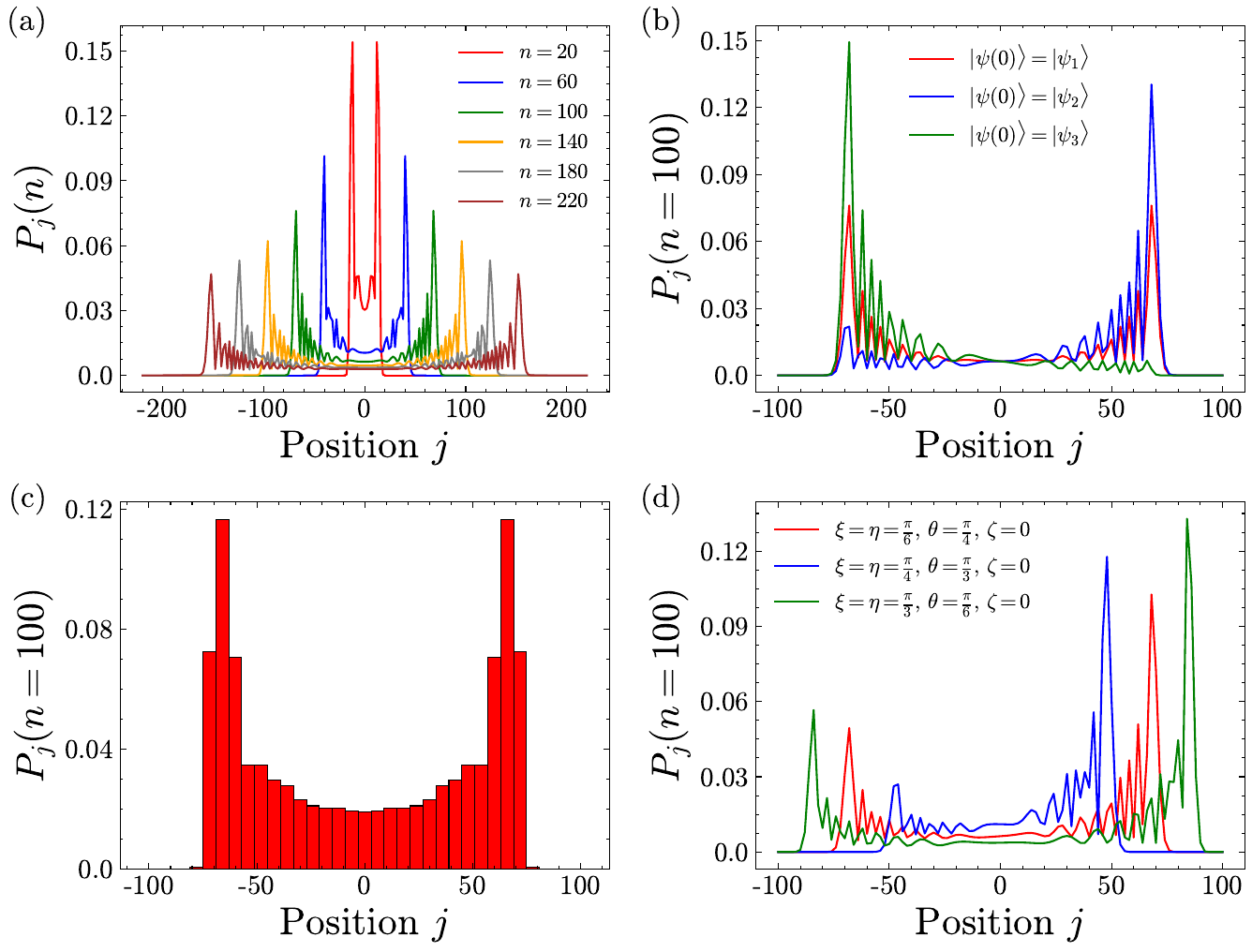}
\caption{\em The sensitivity of the position probability distribution $P_j(n)$ of the quantum walk to changes in the input parameters. (a) The $P_j(n)$ of multiple Hadamard walks with varying number of time steps $n$ and initial condition given by $\vert \psi(0) \rangle = {\frac{1}{\sqrt{2}} (\left| \uparrow \right\rangle + i \left| \downarrow \right\rangle ) \otimes \vert 0 \rangle}$. (b) The $P_j(n=100)$ of Hadamard walks with initial conditions $\vert \psi_1 \rangle = {\frac{1}{\sqrt{2}} (\left| \uparrow \right\rangle + i \left| \downarrow \right\rangle ) \otimes \vert 0 \rangle}$, $\vert \psi_2 \rangle = {\left| \uparrow \right\rangle \otimes \vert 0 \rangle}$, and $\vert \psi_3 \rangle = { ( \frac{-i}{2} \left| \uparrow \right\rangle + \frac{i\sqrt{3}}{2} \left| \downarrow \right\rangle ) \otimes \vert 0 \rangle}$. (c) Histogram of the $P_j(n=100)$ of a Hadamard walk with initial condition $\vert \psi(0) \rangle = {\frac{1}{\sqrt{2}} (\left| \uparrow \right\rangle + i \left| \downarrow \right\rangle ) \otimes \vert 0 \rangle}$. (d) The $P_j(n=100)$ of multiple quantum walks with different $U_{\xi,\theta,\zeta}$ coins (Eq.~(\ref{SU2coinDefinition})) and initial condition given by $\vert \psi(0) \rangle = {\frac{1}{\sqrt{2}} (\left| \uparrow \right\rangle + i \left| \downarrow \right\rangle) \otimes \vert 0 \rangle}$.} 
\label{FigureProbabilityDistributions}
\end{center}
\end{figure}

In Fig.~\ref{FigureProbabilityDistributions}(a), the dependence of the probability distribution $P_j(n)$ on the number of time steps $n$ is displayed. We utilize the Hadamard coin and the initial condition
\begin{equation}  \label{EqSymmetricInitial}
\vert \psi(0) \rangle =  \left( \frac{1}{\sqrt{2}} \left| \uparrow \right\rangle + \frac{i}{\sqrt{2}} \left| \downarrow \right\rangle \right) \otimes \vert 0 \rangle~, 
\end{equation}
which leads to a symmetric probability distribution. The choice for this specific initial condition and quantum coin will be clarified in Sections~\ref{Sec:initialCondition} and \ref{Sec:quantumcoin}. The resulting distributions $P_j(n)$ are bimodal, characterized by two distinct peaks attributed to interference effects. The peaks move further away from each other as time progresses~\citep{romanelli2004quantum}. The overall bimodal shape of the probability distribution remains roughly consistent with increasing $n$. The increasing width of the distribution with $n$ can be adjusted for by rescaling the positions relative to the maximal position for a given $n$. A methodology based on a representative number of time steps $n$ can thus capture the relevant features of $P_j(n)$.

The distributions in Fig.~\ref{FigureProbabilityDistributions}(a) exhibit a bimodal shape, which is rarely observed in short-term financial return distributions (which tend to be unimodal) but can be useful for modeling investor projections~\citep{Orrell:MoneyMagic, orrell2020quantummathsbook, orrell2021quantumwalkArticle}. In Sections~\ref{Sec:initialCondition} and \ref{Sec:quantumcoin}, however, we show the possibility to compute highly asymmetrical return distributions with a substantially less pronounced bimodal shape. Moreover, we will introduce decoherence effects into the quantum walk algorithm in Section~\ref{Sec:Decoherence}, which also allows one to generate unimodal distributions.

Upon using GBM, the standard deviation of the resulting return distribution grows with the square root of the elapsed time $t$. We already addressed the inclusion of the function $f(t)$ to account for potential undesired anomalous diffusion. However, the idea that the standard deviation grows with $t^{1/2}$ is solely based on the assumption of an underlying Wiener process in the asset price evolution. The function $f(t)$ in Eq.~(\ref{EquationQuantumDiffusionForStocks}) can also account for potential ambiguities in this scaling relation, for example during a financial crash~\citep{sornette2018can}.  

\subsubsection{Dependence on the initial condition}   \label{Sec:initialCondition}
In Fig.~\ref{FigureProbabilityDistributions}(b), the dependence of $P_j(n)$ on the initial condition $\vert \psi(0) \rangle$ is examined. The quantum walk is initialized at position $j=0$, but the initial coin state is varied. The quantum coin is the Hadamard coin, and $n$ is set equal to 100. For the initial condition $\vert \psi(0) \rangle = {\frac{1}{\sqrt{2}} (\left| \uparrow \right\rangle + i \left| \downarrow \right\rangle ) \otimes \vert 0 \rangle}$, a fully symmetric distribution is retrieved. It could be argued that for a fair price of the asset, the distribution of price increments should be symmetric, resulting in an equal amount of sellers and buyers~\citep{orrell2020quantummathsbook}. However, selectively tuning the initial condition of the quantum walk allows one to bias the evolution towards increasing or decreasing asset prices (Fig.~\ref{FigureProbabilityDistributions}(b)). This offers a useful tool to model economic bubbles, bubble bursts or quick market corrections that often succeed a bubble, stock market crashes, and other trends, e.g.~a bear or bull market. For example, the initial condition $\vert \psi(0) \rangle =  \left| \uparrow \right\rangle \otimes \vert 0 \rangle$ gives rise to destructive interference for paths with negative increments. The resulting probability distribution can be associated with a bubble, reflected in an escalation of the market value of financial assets. These interference effects, that originate from the discrete-time quantum walk, provide a novel and valuable framework to address uncertainty in the temporal evolution of asset prices, and to effectively propagate the uncertainty over time. This inclusion of interference effects is reminiscent of how they have been integrated into quantum cognition theory.

The level of detail in the distinctive peaks in the probability distribution arising from interference effects in the quantum walk can be mitigated through histogram aggregation, as shown in Fig.~\ref{FigureProbabilityDistributions}(c). By applying this smoothening procedure, we have removed this spiky behavior that is not expected to be observed in actual financial data.

\subsubsection{Dependence on the quantum coin parameters (\texorpdfstring{$\eta = \xi - \zeta$}{(eta = xi - zeta)}, \texorpdfstring{$\theta$)}{theta}}    \label{Sec:quantumcoin}
The Cayley-Klein parameters ($\eta = \xi - \zeta$, $\theta$) in Eq.~(\ref{SU2coinDefinition}) are a selective instrument to set biases towards either increasing or decreasing asset prices. Fig.~\ref{FigureProbabilityDistributions}(d) shows the probability distribution $P_j(n=100)$ of three quantum walks with a $U_{\xi,\theta,\zeta}$ quantum coin and the symmetric initial condition (Eq.~(\ref{EqSymmetricInitial})). The conclusions drawn from the examination of various initial conditions concerning the biases can be extended to this context, where one can introduce an even larger variety of market tendencies by tuning the parameters $\eta$ and $\theta$. By examining the impact of the quantum walk's initial condition and the Cayley-Klein parameters, we can evaluate its effectiveness for modeling the dynamics of extreme market conditions.

\begin{figure}[thb]
\begin{center}
\includegraphics[width=0.5\textwidth,height=2cm,keepaspectratio,angle=0,scale=4.5]{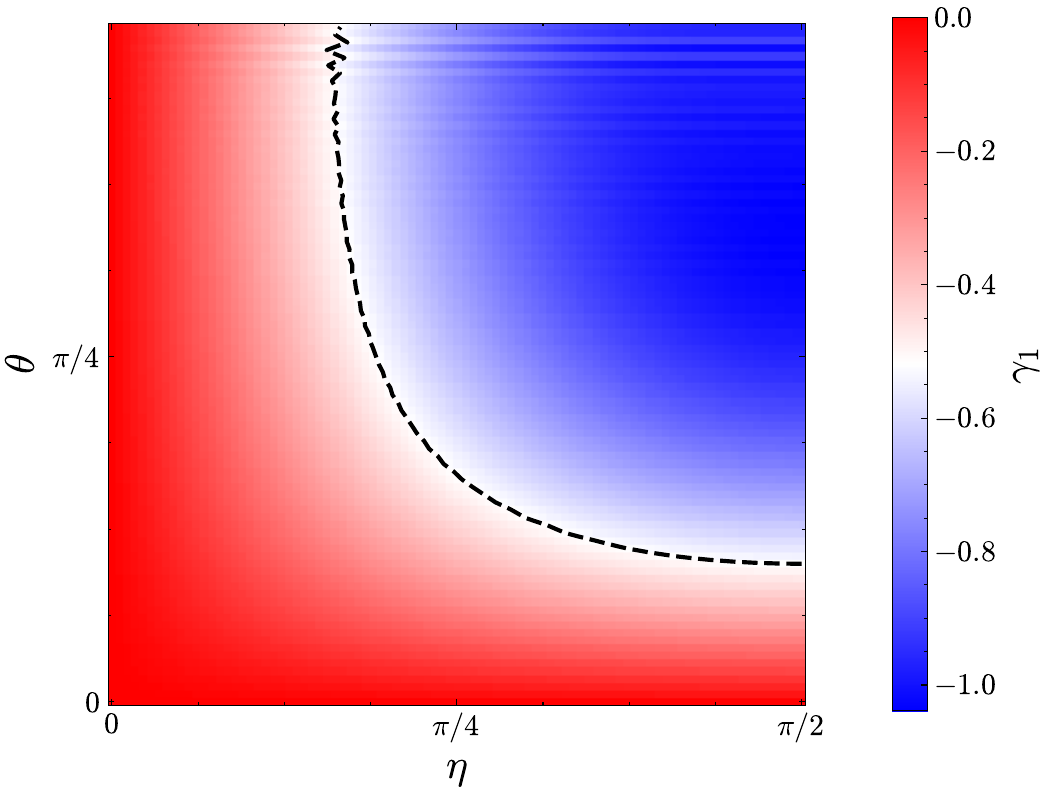}
\caption{\em Heat map with the $(\eta , \theta)$-dependence of the skewness $\gamma_1$ of the position probability distribution $P_j(n=100)$ resulting from a quantum walk with the $U_{\xi,\theta,\zeta}$ coin (Eq.~(\ref{SU2coinDefinition})). The initial condition of the quantum walks is $\vert \psi(0) \rangle = {\frac{1}{\sqrt{2}} (\left| \uparrow \right\rangle + i \left| \downarrow \right\rangle ) \otimes \vert 0 \rangle}$. The limit $\lim_{\theta \to \pi/2} \gamma_1 = \frac{0}{0}$ and leads to numerical instabilities (not shown). The dashed line separates regions with $-1.04 \leq \gamma_1 \leq -0.52$ and with $-0.52 \leq \gamma_1 \leq 0$.}
\label{FigureHeatmapSkewness}
\end{center}
\end{figure}

The asymmetry of a probability distribution can be quantified by the skewness
\begin{equation}
    \gamma_1 = \frac{\kappa_3}{\kappa_2^{3/2}}~,
\end{equation}
where $\kappa_i$ is the $i$th cumulant of the distribution. In Fig.~\ref{FigureHeatmapSkewness}, the skewness of the probability distribution $P_j(n=100)$ of the quantum walk with the symmetric initial condition of Eq.~(\ref{EqSymmetricInitial}) and the $U_{\xi,\theta,\zeta}$ quantum coin is investigated as a function of $\eta$ and $\theta$. For this symmetric initial condition, $\eta$ and $\theta$ are confined to the interval $[0, \pi/2]$, since all other parameter values yield distributions equivalent to those obtained within the restricted interval (modulo a reflection about $j=0$ in some situations). For these $(\eta , \theta)$-combinations, the skewness adopts negative values, indicating a left-skewed distribution that leans to the right, as can be seen in Fig.~\ref{FigureProbabilityDistributions}(d). This can be interpreted as a bias towards positive increments.

\begin{figure}[thb]
\begin{center}
\includegraphics[width=0.5\textwidth,height=2cm,keepaspectratio,angle=0,scale=4.5]{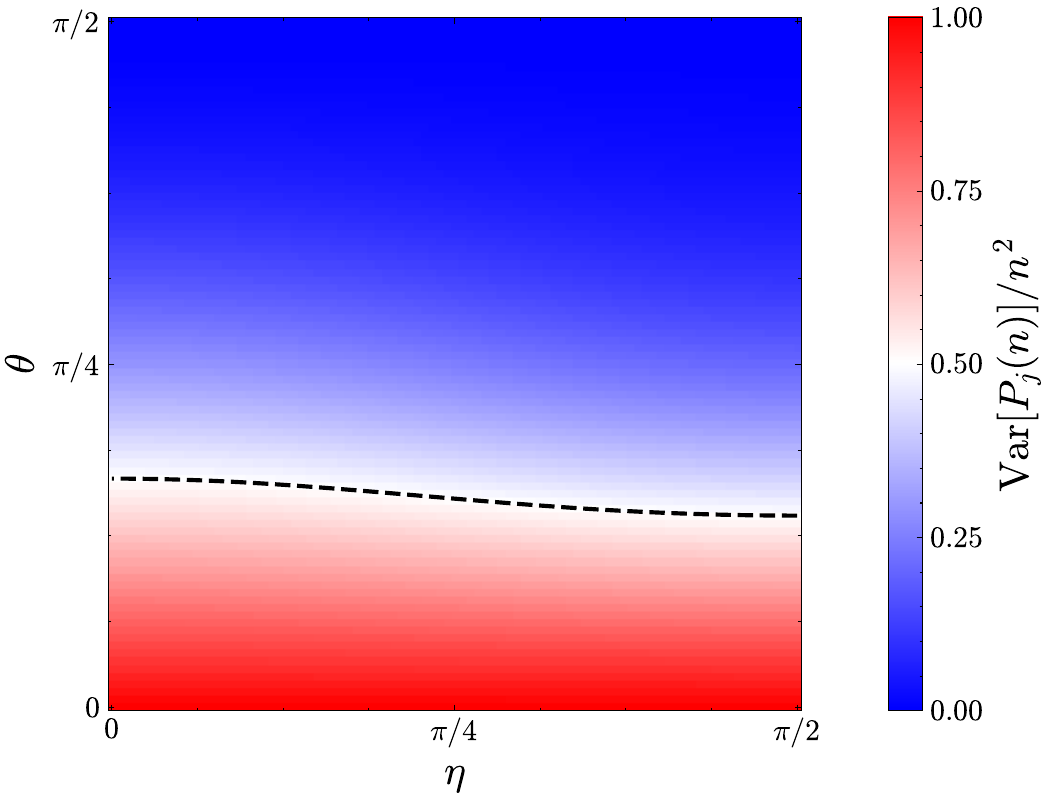}
\caption{\em Heat map with the $(\eta , \theta)$-dependence of the variance of $P_j(n=100)$ resulting from a quantum walk with the $U_{\xi,\theta,\zeta}$ coin relative to the square of the number of time steps $n$ ($\text{Var}[P_j(n)]/n^2$ with $n=100$). The initial condition is $\vert \psi(0) \rangle = {\frac{1}{\sqrt{2}} (\left| \uparrow \right\rangle + i \left| \downarrow \right\rangle ) \otimes \vert 0 \rangle}$. The dashed line separates regions with $0 \leq \text{Var}[P_j(n)]/n^2 \leq 0.5$ and with ${0.5 \leq \text{Var}[P_j(n)]/n^2 \leq 1}$.}
\label{FigureHeatMap}
\end{center}
\end{figure}

When modeling asset price time series, an essential consideration is the analysis of the diffusion properties in the temporal evolution. The diffusion properties of financial products are of extreme relevance when evaluating risk. In Fig.~\ref{FigureHeatMap}, the variance of the probability distribution $P_j(n=100)$ of the quantum walk with the symmetric initial condition of Eq.~(\ref{EqSymmetricInitial}) and the $U_{\xi,\theta,\zeta}$ quantum coin (Eq.~(\ref{SU2coinDefinition})) is investigated as a function of $\eta$ and $\theta$. Since the variance grows quadratically with the number of time steps $n$ in the unitary discrete-time quantum walk, the variance is divided by $n^2$. The strongest level of dependence lies in the parameter $\theta$. For a $U_{\theta}$ coin (Eq.~(\ref{UthetacoinDefinition})), the scaling behavior of the variance can be well approximated by~\citep{chandrashekar2008optimizing}
\begin{equation}    \label{EqVarianceForUTheta}
    \underset{j}{\textrm{Var}}[P_j(n)] / n^2 \approx 1 - \sin{\theta}~.
\end{equation}

The variance provides information about the deviation of the distribution from its mean value, and can be associated with the unpredictability of the outcome upon executing an experiment. If the variance adopts large values ($\theta \lesssim \pi / 3 $), the outcome can be difficult to predict. For small variances ($\theta \gtrsim \pi / 3 $), there may be less uncertainty about the possible outcomes.

\begin{figure}[thb]
\begin{center}
\includegraphics[width=0.5\textwidth,height=2cm,keepaspectratio,angle=0,scale=4.5]{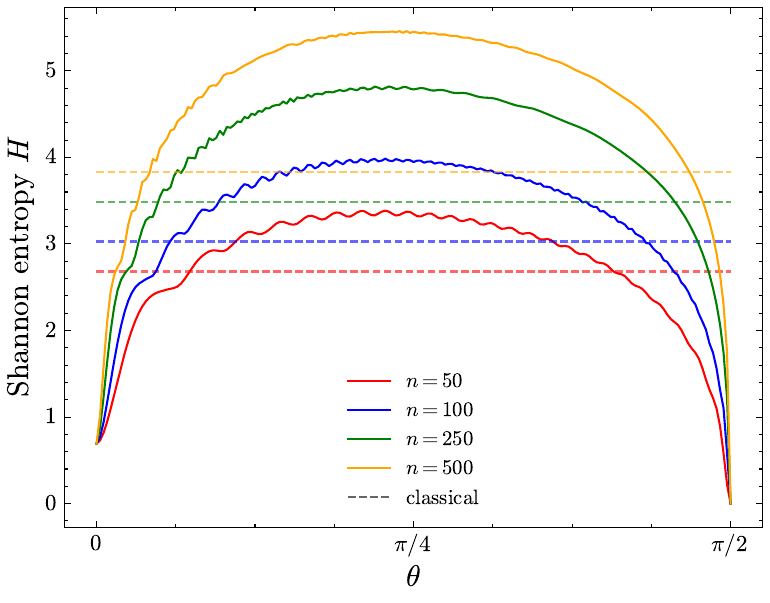}
\caption{\em The Shannon entropy of $P_j(n)$ for a quantum walk with a $U_{\theta}$ coin of Eq.~(\ref{UthetacoinDefinition}) for various values of $n$ (solid lines). The initial condition is $\vert \psi(0) \rangle = {\frac{1}{\sqrt{2}} (\left| \uparrow \right\rangle + i \left| \downarrow \right\rangle ) \otimes \vert 0 \rangle}$. For the sake of reference, we display the Shannon entropy of the position probability distribution resulting from a classical random walk with an equal number of time steps $n$ (dashed lines).}
\label{FigureShannonEntropy}
\end{center}
\end{figure}

The Shannon entropy $H(n)$ can also be linked with the delocalization of $P_j(n)$ and can be used as a measure to quantify the uncertainty associated with $P_j(n)$~\citep{pires2020quantum, ebrahimi1999ordering}
\begin{equation}    \label{EqShannonEntropy}
    H(n) = -\sum_{j=-n }^{+n} P_j(n) \log_e{P_j(n)}~.
\end{equation}
Fig.~\ref{FigureShannonEntropy} shows that the probability distributions from quantum walks with a $U_{\theta}$ coin (Eq.~(\ref{UthetacoinDefinition})) have a larger degree of complexity over a large range of $\theta$-values than the distribution resulting from the corresponding classical random walk. The entropy reaches its maximum for $\theta \approx \pi/4$, which corresponds with the Hadamard coin. When applying the Hadamard coin, the associated probability distribution carries the most uncertainty about the possible outcome of all $U_{\theta}$ coins. In financial terms, the Hadamard coin can capture the largest variety of outcomes in the asset price evolution. The entropy $H(n)$ is not symmetric about $\theta = \pi/4$. For small values of $\theta$, the $U_{\theta}$ resembles the $\sigma_{z}$ Pauli matrix, with the walker ending up near $j=-n$ and $j=+n$ after the operation of $\widehat{V}^n$. This results in $H(n) = \log_e{2}$ for $\theta = 0$. For $\theta$ approaching $\pi/2$, the quantum coin resembles the $\sigma_{x}$ Pauli matrix. Then the $P_j(n)$ is highly concentrated around the origin, resulting in $H(n) = 0$ for $\theta = \pi/2$.

\subsection{Decoherence}    \label{Sec:Decoherence}
The terminology ``decoherence'' refers to the loss of quantum coherence, and establishes connections between classical and quantum behavior~\citep{gamble2009demystifying}. Decoherence arises from the fact that quantum systems are seldom completely isolated. Interactions with the environment, for example in a measurement process, give rise to loss of coherence since the system becomes entangled with a large number of environmental degrees of freedom~\citep{schlosshauer2019quantum}. Decoherence involves information leakage into the environment, an integral aspect when modeling the temporal evolution of any phenomenon which underscores the link between physics and information theory.

There are multiple methodologies to introduce decoherence in the quantum walk algorithm~\citep{ishak2021entropy, romanelli2005decoherence, romanelli2011coinflipping, mackay2002quantum, brun2003quantumDecoherentCoins, brun2003quantumManyCoins, jayakody2018transfiguration}. For the current purposes, we prefer methodologies with a straightforward interpretation and implementation.

\begin{figure}[ht]
\centering

\begin{tikzpicture}[>=Stealth, thick]

\foreach \x/\label in {-6/$j-1$,-3/$j$,0/$j+1$,3/$j+2$} {
    \draw (\x,-1.5) -- (\x,1.5) node[below] at (\x,-1.5) {\label};
}

\draw (0,0.5) -- (3.5,0.5) node[right] {$\left| \uparrow \right\rangle$};
\draw[dotted] (-3,0.5) -- (0,0.5);
\draw (-6.5,0.5) -- (-3,0.5);
\draw (0,-0.5) -- (3.5,-0.5) node[right] {$\left| \downarrow \right\rangle$};
\draw[dotted] (-3,-0.5) -- (0,-0.5);
\draw (-6.5,-0.5) -- (-3,-0.5);

\draw[->,thick, blue] (-3,-0.5) to[out=-120,in=-60] (-6,-0.5);
\draw[->,thick, red] (0,-0.5) to[out=120,in=-120] (0,0.5);
\draw[->,thick, blue] (3,-0.5) to[out=-120,in=-60] (0,-0.5);
\draw[->,thick, blue] (-6,0.5) to[out=60,in=120] node[midway, above] {$a_{j}(n+1)$} (-3,0.5);
\draw[->,thick, red] (-3,0.5) to[out=-60,in=60] node[midway, right] {$b_{j}(n+1)$} (-3,-0.5);
\draw[->,thick, blue] (0,0.5) to[out=60,in=120] (3,0.5);

\node[align=center] at (-7.5, 0) {(b)};

\begin{scope}[yshift=4cm]
\foreach \x/\label in {-6/$j-1$,-3/$j$,0/$j+1$,3/$j+2$} {
    \draw (\x,-1.5) -- (\x,1.5) node[below] at (\x,-1.5) {\label};
}

\foreach \y/\ket in {-0.5/{$\left| \downarrow \right\rangle$},0.5/{$\left| \uparrow \right\rangle$}} {
    \draw (-6.5,\y) -- (3.5,\y) node[right] {\ket};
}

\draw[->,thick, blue] (-3,-0.5) to[out=-120,in=-60] (-6,-0.5);
\draw[->,thick, blue] (0,-0.5) to[out=-120,in=-60] node[midway, below] {$b_{j}(n+1)$} (-3,-0.5);
\draw[->,thick, blue] (3,-0.5) to[out=-120,in=-60] (0,-0.5);
\draw[->,thick, blue] (-6,0.5) to[out=60,in=120] node[midway, above] {$a_{j}(n+1)$} (-3,0.5);
\draw[->,thick, blue] (-3,0.5) to[out=60,in=120] (0,0.5);
\draw[->,thick, blue] (0,0.5) to[out=60,in=120] (3,0.5);
\node[align=center] at (-7.5, 0) {(a)};
\end{scope}

\end{tikzpicture}

\caption{\em A schematic illustration of the effect of a disrupted link on the conditional translation (Eq.~(\ref{DefinitionTranslation})). The upper panel (a) corresponds with no broken links (unitary quantum walk). After applying the quantum coin toss operator, the $\left| \uparrow \right\rangle$-component $a_j(n+1)$ is shifted to the right in the transition from time step $n$ to $n+1$, and the $\left| \downarrow \right\rangle$-component $b_j(n+1)$ is shifted to the left. In the lower panel (b), the link between sites $j$ and $j+1$ is broken, and the flux that is carried over this link in an undisrupted quantum walk is diverted.}
\label{FigureIllustrationBrokenLinks}
\end{figure}
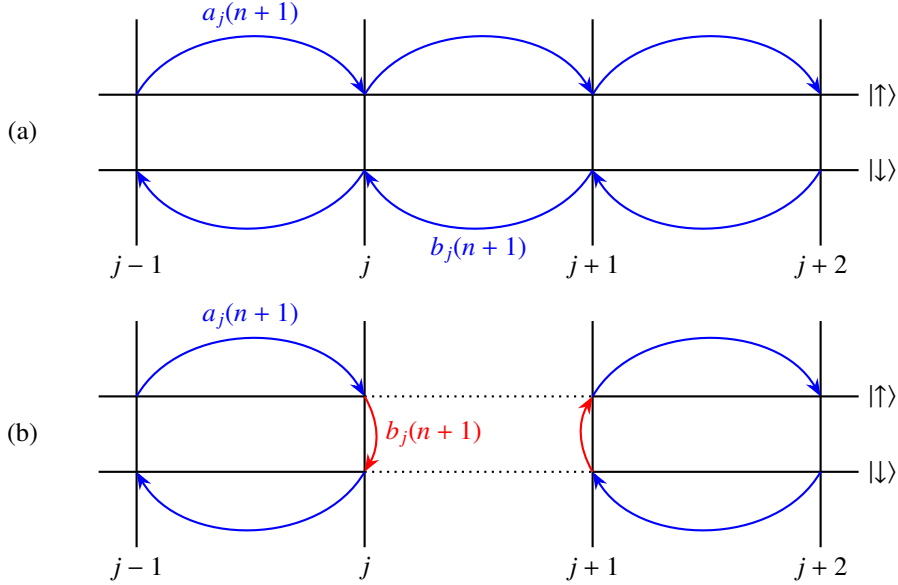

A mechanism to include decoherence consists of randomly breaking links between sites in the one-dimensional grid of the quantum walk~\citep{romanelli2005decoherence, romanelli2011coinflipping}. At each time step, a link has a probability $p$ of being disrupted. For a site with no disrupted links, the quantum coin toss and the conditional shift operator are applied as usual. If one or both links at position $j$ are broken, translation across an opened link becomes impossible, which modifies the temporal evolution. For a broken link at the right of site $j$, the upper component of the coin space (corresponding with $\left| \uparrow \right\rangle$) at position $j$ still receives probability flux from site $j-1$. However, the outgoing flux of the upper component $\left| \uparrow \right\rangle$ at site $j$ gets diverted to the lower component $\left| \downarrow \right\rangle$ at the same site in order to preserve the flux (Fig.~\ref{FigureIllustrationBrokenLinks}). For the $U_{\theta}$ coin of Eq.~(\ref{UthetacoinDefinition}), the update of the amplitudes for $n \to n + 1$ involves the algorithm
\begin{equation}
\label{Equation:DecoherenceRightLink}
\left\{
\begin{array}{lclcl}
    a_j(n+1) = \cos{\theta} ~ a_{j-1}(n) + \sin{\theta} ~ b_{j-1}(n)  \\
    b_j(n+1) = \cos{\theta} ~ a_{j}(n) +  \sin{\theta} ~ b_{j}(n) ~.
\end{array}
\right.
\end{equation}

For a broken link on the left of site $j$, namely the $(j-1, j)$-link, we have
\begin{equation}
\label{Equation:DecoherenceLeftLink}
\left\{
\begin{array}{lclcl}
    a_j(n+1) = \sin{\theta} ~ a_{j}(n) -  \cos{\theta} ~ b_{j}(n) \\
    b_j(n+1) = \sin{\theta} ~ a_{j+1}(n) - \cos{\theta} ~ b_{j+1}(n)~.
\end{array}
\right.
\end{equation}

If both links at site $j$ are disrupted, namely the $(j-1, j)$- and $(j, j+1)$-links, the upper and lower component of the coin space are exchanged, leading to
\begin{equation}
\label{Equation:DecoherenceBothLinks}
\left\{
\begin{array}{lclcl}
    a_j(n+1) = \sin{\theta} ~ a_{j}(n) - \cos{\theta} ~ b_{j}(n)  \\
    b_j(n+1) = \cos{\theta} ~ a_{j}(n) +  \sin{\theta} ~ b_{j}(n)~.
\end{array}
\right.
\end{equation}

\begin{figure}[thb]
\begin{center}
\includegraphics[width=0.5\textwidth,height=2cm,keepaspectratio,angle=0,scale=4.5]{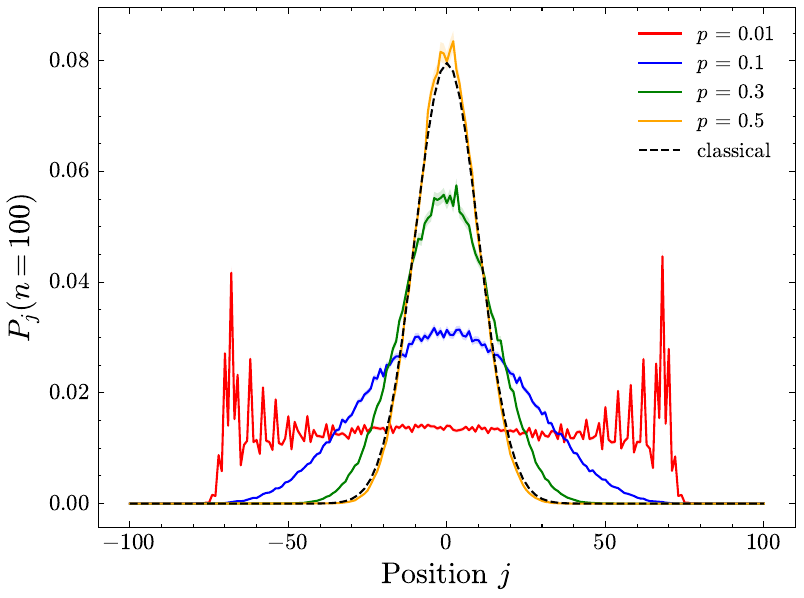}
\caption{\em The $P_j(n=100)$ of multiple Hadamard walks with decoherence through the stochastic disruption of links. The initial condition is given by $\vert \psi(0) \rangle = {\frac{1}{\sqrt{2}} (\left| \uparrow \right\rangle + i \left| \downarrow \right\rangle ) \otimes \vert 0 \rangle}$. For each value of the parameter $p$, the displayed $P_j(n)$ are the average over 1000 simulations. The $P_j(n)$ are normalized by requiring that the integral over the positions has the same value as for the distribution resulting from a classical random walk with an equal number of time steps $n$ (dashed line). The lightly colored zones around the graphs represent the standard deviation of the mean over the 1000 simulations, which are nearly insignificant in comparison to the linewidth of the graphs.}
\label{FigureDecoherenceBrokenLinks}
\end{center}
\end{figure}

Fig.~\ref{FigureDecoherenceBrokenLinks} shows the position probability distributions $P_j(n)$ resulting from multiple Hadamard walks with various values of the probability $p$ to disrupt links between positions. For $p=0.01$, the distribution resembles the outcome of the unitary Hadamard walk. For increasing values of $p$, we see the characteristic transition from quantum to classical behavior, which is completed for $p=0.5$.

An alternative and intuitive way of introducing decoherence in the unitary quantum walk is by utilizing the coin 
\begin{equation}   \label{EqDecoherenceCoin}
    U_{\xi=0, \theta, \zeta} =  \begin{pmatrix}
   \cos{\theta} & e^{i\zeta} \sin{\theta} \\ 
    e^{-i\zeta} \sin{\theta} & - \cos{\theta}
\end{pmatrix}~,
\end{equation}
and drawing $\zeta$ from a uniform distribution over the interval $\left[0, 2 \pi \right[$ with a probability $\Tilde{p}$ at each time step~\citep{mackay2002quantum}. With a probability $1 - \Tilde{p}$, $\zeta$ is set equal to 0. This methodology introduces random phases, which destroy the interference between the different paths in the unitary quantum walk, eventually eliminating the quantum coherence. We have checked that this methodology produces results that are similar to those presented in Fig.~\ref{FigureDecoherenceBrokenLinks}. For this methodology, the classical situation is reached for $\Tilde{p}=1$. 

Other intuitive mechanisms to introduce decoherence consist of performing a measurement of the coin state with a certain probability at each time step~\citep{brun2003quantumDecoherentCoins}, introducing a well-chosen type of complete positive map defined on the coin degree of freedom~\citep{jayakody2018transfiguration}, periodically performing a joint measurement of the position and coin state~\citep{romanelli2005decoherence}, using a different quantum coin at each time step~\citep{brun2003quantumManyCoins}, and introducing a bit-flip quantum noise channel causing random flips of the coin state with a certain probability~\citep{ishak2021entropy}.

Fig.~\ref{FigureDecoherenceBrokenLinks} shows that decoherence brings the distributions from a quantum walk closer to those of a classical random walk. For a full transition to classical behavior, the standard deviation of the distribution no longer scales with $n$, but rather with $n^{1/2}$ as in the case of GBM. In this context, the model for the asset price evolution based on the discrete-time quantum walk (Eq.~(\ref{EquationQuantumDiffusionForStocks})) can be looked upon as an extension of the model based on GBM (Eq.~(\ref{GeometricBrownianMotion})).

The decoherence effects simulate the impacts stemming from unpredictable and disrupting information in a supposedly perfectly informed market, and tend to destroy the coherence in the quantum diffusive process. The methodologies designed for introducing decoherence rely on incorporating some sort of randomness into the system (e.g.~randomly breaking links with a given probability, introducing a random phase in the quantum coin), which is reminiscent of the random increments in GBM.

In the approaches based on the random disruption of links and the introduction of a random phase, there is a tunable parameter that measures the degree to which the quantum walk has collapsed to a classical diffusive process. In Fig.~\ref{FigureDecoherenceBrokenLinks}, the distributions for $p=0.1$ and $p=0.3$ are non-Gaussian and exhibit fat tails, yielding a higher probability of more extreme events than the classical model. This can be useful to overcome issues encountered in classical models lacking this property (e.g.~GBM). 

\begin{figure}[thb]
\begin{center}
\includegraphics[width=0.5\textwidth,height=2cm,keepaspectratio,angle=0,scale=4.5]{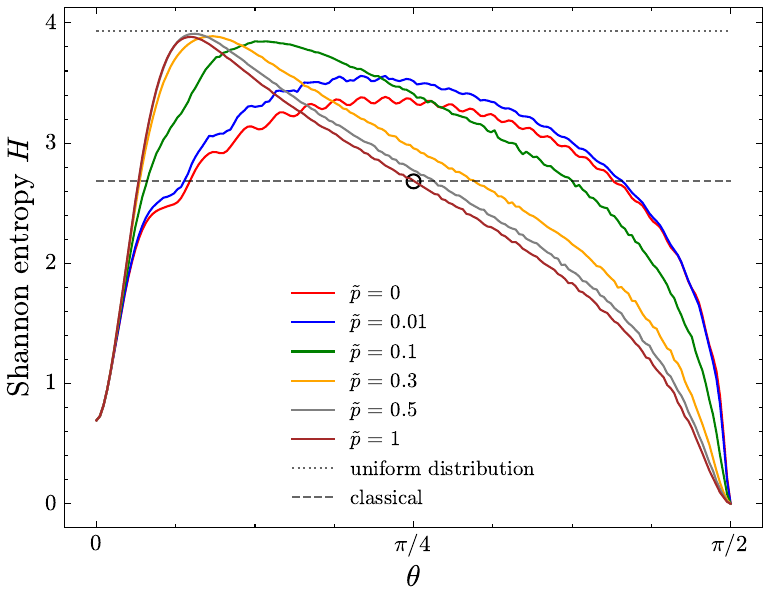}
\caption{\em The Shannon entropy of $P_j(n=50)$ for a quantum walk with the coin of Eq.~(\ref{EqDecoherenceCoin}) (solid lines). The impact of decoherence is shown for various choices of $\Tilde{p}$. The initial condition is $\vert \psi(0) \rangle = {\frac{1}{\sqrt{2}} (\left| \uparrow \right\rangle + i \left| \downarrow \right\rangle ) \otimes \vert 0 \rangle}$. The distributions $P_j(n=50)$ are averaged over 1000 simulations. The normalization is guaranteed by requiring that the discrete probabilities add up to one. For the sake of reference, we display the Shannon entropy of a uniform distribution (dotted line) and the position probability distribution resulting from a classical random walk with 50 time steps (dashed line).}
\label{FigureEntropyAndDecoherence}
\end{center}
\end{figure}

To conclude this subsection, we use the methodology based on the introduction of a random phase (Eq.~(\ref{EqDecoherenceCoin})) to investigate the influence of decoherence effects on the Shannon entropy (Eq.~(\ref{EqShannonEntropy})) of the probability distribution resulting from a quantum walk (Fig.~\ref{FigureEntropyAndDecoherence}). For very low values of the probability $\Tilde{p}$ to introduce a random phase in the quantum coin (e.g.~$\Tilde{p}=0.01$), the results closely resemble those of the unitary quantum walk, although having a slightly higher value of the Shannon entropy over the entire $\theta$-range. For higher values of the parameter $\Tilde{p}$, the Shannon entropy attains its maximum value for $\pi/16 \lesssim \theta \lesssim \pi / 8$. Under these conditions, the distribution is very wide, and $H(n)$ approaches the value for a uniform distribution. For $\Tilde{p}=1$ and $\theta = \pi/4$, the Hadamard quantum walk has collapsed to the classical situation, and the entropy becomes equal to the one for the probability distribution resulting from the classical random walk (open circle in Fig.~\ref{FigureEntropyAndDecoherence}). 

As mentioned in Section \ref{Sec:quantumcoin}, the Shannon entropy is a relevant measure to investigate in a financial context, since it can provide insights into market volatility and stability~\citep{sheraz2015entropy}. Indeed, a higher entropy suggests greater uncertainty or variability in the return distribution (for example for $\theta \approx \pi/16$ in Fig.~\ref{FigureEntropyAndDecoherence}). Moreover, it possesses the potential to discriminate between different market regimes. For example, a notable increase in entropy could signal the onset of a turbulent market period or the occurrence of an unforeseen event~\citep{drzazga2023entropy}.

\subsection{Advantages over classical methodologies}
Table \ref{TableQuantumAdvantages} lists some selected restrictions of classical methodologies for modeling financial return distributions. The second column highlights how the quantum walk model can overcome these restrictions. 

\begin{table}
\begin{tabular}{ | p{6cm}|p{6cm}| }
\hline
\textbf{Restriction of classical model} & \textbf{Cure provided by quantum walk model} \\
\hline
GBM: inability to capture large price changes & great flexibility in shape of return distribution (bimodal, bias towards increasing or decreasing prices, occurrence of extreme changes, inclusion of decoherence effects) \\
\cline{1-2}
GBM: symmetric return distributions & biases towards increasing or decreasing prices by tuning ($\eta$, $\theta$) and $\vert \psi (0) \rangle$ \\ 
\cline{1-2}
GBM: Markovian process, i.i.d.~process, randomness (EMH assumption) & non-Markovian properties due to interference effects \\
\hline
LS\&PL: divergent higher moments & higher moments do not diverge \\
\cline{1-2}
LS\&PL: restricted to morphological fit to return distributions & quantum walk model produces return distributions through its time evolution \\
\hline
\end{tabular}
\caption{\label{TableQuantumAdvantages}Selected limitations of the discussed classical models (geometric Brownian motion (GBM) and Lévy stable and power law distributions (LS\&PL)) and how the quantum walk model can overcome them.}
\end{table}

In Section~\ref{Sec:quantumcoin}, we have illustrated that the quantum walk model can be tuned to result in return distributions with a specific level of asymmetry and dispersion, as quantified respectively by the skewness (Fig.~\ref{FigureHeatmapSkewness}) and the variance (Fig.~\ref{FigureHeatMap}). The clearly distinct and well-resolved regions in the parameter space can be associated with a rich variety of distinct market behaviors in asset price evolution. For instance, high values of the variance indicate increased market uncertainty, while an asymmetric distribution indicates a bias towards either increasing or decreasing prices. Furthermore, by analyzing the Shannon entropy, we have demonstrated that the quantum walk model can generate return distributions with a higher level of complexity than classical random walks, as evidenced by increased delocalization (Figs.~\ref{FigureShannonEntropy} and \ref{FigureEntropyAndDecoherence}).

\begin{figure}[thb]
\begin{center}
\includegraphics[width=0.5\textwidth,height=2cm,keepaspectratio,angle=0,scale=4.5]{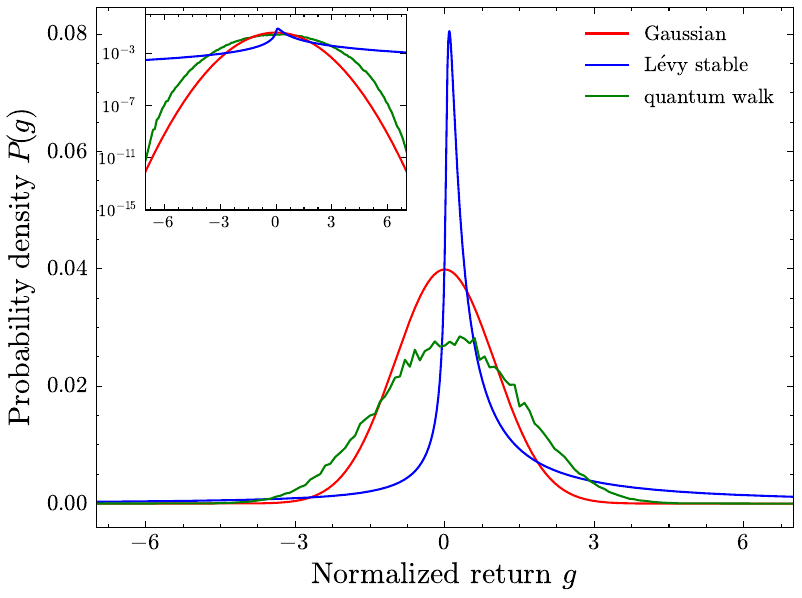}
\caption{\em Prototypical examples of return distributions from three models: (i) a Gaussian distribution ($\mu = 0$, $\sigma = 1$); (ii) a Lévy stable distribution ($\alpha = \beta = 0.5$, $\mu=0$, $c = 1/\sqrt{2}$); and (iii) a quantum walk with decoherence by randomly disrupting links ($n = 100$, $\vert \psi(0) \rangle = \left| \uparrow \right\rangle \otimes \vert 0 \rangle$, Hadamard coin, $p=0.3$, averaged over 1000 simulations). The inset figure displays the same distributions with a logarithmic vertical scale.}
\label{FigureReturnDistributions}
\end{center}
\end{figure}

Fig.~\ref{FigureReturnDistributions} displays three prototypical examples of return distributions, which are computed by using: (i) a Gaussian distribution, (ii) a Lévy stable distribution, and (iii) a quantum walk with decoherence through the stochastic disruption of links. The normalized returns are defined as in Eq.~(\ref{EqDefinitionNormalizedReturn}), where the denominator is given by the standard deviation of the Gaussian distribution. Tuning the parameters of the Lévy stable distribution allows one to generate an asymmetric distribution with a more leptokurtic shape than a Gaussian distribution. The initial condition of the quantum walk is given by $\vert \psi (0) \rangle = \left| \uparrow \right\rangle \otimes \vert 0 \rangle$, which models a bias towards increasing asset prices. As previously mentioned, the return distribution resulting from the quantum walk with decoherence has fatter tails than the Gaussian distribution, leading to a higher probability of extreme events. These results suggest that the quantum walk algorithm offers a promising and flexible tool for modeling return distributions of financial products. 

\section{Conclusions}     \label{Sec:Conclusions}
Classical methodologies have been a powerful tool to represent the temporal evolution of asset prices and to quantify stochastic deviations in financial time series. With the prospect of advanced quantum computation technologies, we have explored the potential of the discrete-time quantum walk for modeling financial time series. The concept of a quantum walk adds the flexibility of superposition and interference, which can be used to model competing driving mechanisms in the underlying dynamics. While the quantum walk model introduces additional parameters, this added complexity is balanced by its ability to model certain financial return characteristics that fall beyond the scope of classical models. In this respect, quantum walks have the potential to enrich our modeling capabilities beyond those of classical approaches.

In this paper, several restrictions of classical methodologies have been pointed out, e.g.~the inability of some models to capture large price changes, the restricted flexibility in generating asymmetric distributions, the assumption of no discernible pattern in price fluctuations, the occurrence of diverging higher moments, and the limited insight into the dynamics that drive the price evolution. We illustrated the potential of the quantum walk algorithm to overcome these limitations, as well as the possibility to control the underlying dynamics by tuning the parameters of the quantum walk, leading to distributions that can match the returns of financial products in a plethora of circumstances. By examining the diffusion properties across the allowed values of the quantum walk parameters, we identified their impact on the dynamics and the resulting distributions, which is instrumental in studying the diffusion properties of asset prices. Our findings open new opportunities for understanding and predicting market dynamics. The demonstrated flexibility of the quantum walk model in capturing the market tendencies reflected in return distributions shows the potential for financial modeling, and can offer new insights into risk assessment. 

An important line for future research lies in the application of the model to historic financial data, in particular to return distributions. This will further deepen our understanding of the financial interpretation of the parameters governing the quantum walk dynamics. In this respect, it can also be further investigated to which exact economic scenarios the model is best applicable. Ongoing research includes an in-depth analysis of historical financial data to refine the model's application and interpretive parameters.

\appendix
\section{Relevant angles in the quantum coin for determining \texorpdfstring{$P_j(n)$}{Pj(n)} for a quantum walk initialized at position \texorpdfstring{$j = 0$}{j = 0}} \label{sec:appendix}

For a quantum walk initialized at position $j = 0$ at time $n=0$, the initial conditions for the coin amplitudes $(a_j(n),b_j(n))$ read
\begin{equation}
\label{EqAppendixInitialCondition}
\left\{
\begin{array}{lcl}
    a_j(0) &= &  \delta_{0j} \;  a_0(0) \\
    b_j(0) &= & \delta_{0j} \; b_0(0) ~,
\end{array}
\right.    
\end{equation}
with $ a_0(0), b_0 (0) \in \mathbb{C}$ obeying the normalization condition $ | a_0(0) | ^2 + | b_0(0) | ^2 = 1$.
 
The unitary operator $\widehat{V}$ of Eq.~(\ref{EqOperotarV}) with the $U_{\xi, \theta, \zeta}$ coin gives rise to the following update rules of the coin amplitudes in the $ n \rightarrow n + 1 $ time step 
\begin{equation}
\label{Equation:Recurrence}
\left\{
\begin{array}{lclcl}
    a_j(n+1) &= &  \left( e^{+i \xi} \cos{\theta} \right) ~ a_{j-1}(n) & + &  \left( e^{+i \zeta} \sin{\theta} \right) ~b_{j-1}(n)  \\
    b_j(n+1) &= & \left( e^{-i \zeta} \sin{\theta} \right) ~ a_{j+1}(n)  & - &  \left( e^{-i \xi} \cos{\theta} \right) ~ b_{j+1}(n) ~,
\end{array}
\right.
\end{equation}
which corresponds with moving probability amplitudes from position $j - 1$  to $j$ and from position $j + 1$ to $j$. We denote these as updates of type $j-1 \rightarrow j$ and $j+1 \rightarrow j$ respectively. Note that there is cross-feeding between the $a_j$ and $b_j$ amplitudes. In order to calculate the $P_j(n)= | a_{j}(n) | ^2 + | b_{j}(n) | ^2$ of Eq.~(\ref{EqprobabilityQuantumWalk}), one iteratively applies the above update rules. The only contributions to $P_j(n)$ arise from the terms $a_0(0) a^\ast_0(0)$, $a_0(0) b^\ast_0(0)$, $b_0(0) a^\ast_0(0)$, and $b_0(0) b^\ast_0(0)$, of which the respective weights are functions of the angles $(\xi, \theta, \zeta)$. Note that an update of the type $j - 1 \rightarrow j$ gives rise to an extra phase factor $e^{+i \xi}$ or $e^{+i \zeta}$ for the $(a_j,b_j)$ amplitudes ($e^{-i \xi}$ or $e^{-i \zeta}$ for the $(a_j^{\ast},b_j^{\ast})$). Similarly, an update of the type $j + 1 \rightarrow j$ gives rise to an extra phase factor $e^{-i \zeta}$ or $e^{-i \xi}$ for the $(a_j,b_j)$ amplitudes ($e^{+i \zeta}$ or $e^{+i \xi}$ for the $(a_j^{\ast},b_j^{\ast})$). Given the initial condition (\ref{EqAppendixInitialCondition}) with only non-vanishing amplitudes at position $j=0$, the $P_j(n)$ is the result of $N_{+} \equiv (n+j)/2$ updates of type $j-1 \rightarrow j$ (total number of steps to the right), and $N_{-} \equiv (n-j)/2$ updates of type $j+1 \rightarrow j$ (total number of steps to the left).

The phase factors in the contributing terms in the expression for $P_j(n)$ that originate from the $N_{+}$ updates of type $j-1 \rightarrow j$, are given by
\begin{equation}
    (e^{i \xi}) ^\lambda (e^{i \zeta}) ^ {N_{+} - \lambda} (e^{-i \xi}) ^\nu(e^{- i \zeta}) ^ {N_{+} - \nu} = e^{i (\xi - \zeta )(\lambda - \nu)}~,
\end{equation}
with $0 \leq \lambda,\nu \in \mathbb{N} \leq N_{+}$. Similarly, the phase factors in the contributing terms in the expression for $P_j(n)$ that originate from the $N_{-}$ updates of type $j+1 \rightarrow j$, are given by
\begin{equation}
    (e^{-i \zeta}) ^\rho (e^{-i \xi}) ^ {N_{-} - \rho} (e^{i \zeta}) ^\tau(e^{i \xi}) ^ {N_{-} - \tau} = e^{i (\xi - \zeta )(\rho - \tau)}~,
\end{equation}
with $0 \leq \rho,\tau \in \mathbb{N} \leq N_{-}$. The values of $\lambda$, $\nu$, $\rho$ and $\tau$ depend on the paths that have their contribution in the quantum walk. It is concluded that for a quantum walk initialized at position $j = 0$, $P_j(n)$ is fully determined by the choices for the angles  $(\eta \equiv \xi - \zeta, \theta) $ in the quantum coin.

\section*{Declaration of Generative AI and AI-assisted technologies in the writing process}
During the preparation of this work, the authors used ChatGPT in order to enhance the readability of this work. After using this tool, the authors reviewed and edited the content as needed and take full responsibility for the content of the publication.

\section*{Declaration of competing interest}
The authors declare that they have no known competing financial interests or personal relationships that could have appeared
to influence the work reported in this paper.

\section*{Acknowledgment}
This project was conducted with the support of the Special Research Fund of Ghent University (projects BOF/DOC/2023/103 and
BOF/BAF/4y/24/1/018).

\section*{Data availability}
No data was used for the research described in the article.

\bibliographystyle{elsarticle-num} 
\bibliography{cas-refs}

\begin{thebibliography}{10}
\expandafter\ifx\csname url\endcsname\relax
  \def\url#1{\texttt{#1}}\fi
\expandafter\ifx\csname urlprefix\endcsname\relax\def\urlprefix{URL }\fi
\expandafter\ifx\csname href\endcsname\relax
  \def\href#1#2{#2} \def\path#1{#1}\fi

\bibitem{Sharma2011}
B.~Sharma, S.~Agrawal, M.~Sharma, D.~Bisen, R.~Sharma, Econophysics: A brief review of historical development, present status and future trends, arXiv preprint arXiv:1108.0977 (2011).

\bibitem{pitowsky2006quantum}
I.~Pitowsky, Quantum mechanics as a theory of probability, in: Physical theory and its interpretation: Essays in honor of Jeffrey Bub, Springer, 2006, pp. 213--240.

\bibitem{busemeyer2015quantum}
J.~R. Busemeyer, Z.~Wang, What is quantum cognition, and how is it applied to psychology?, Current Directions in Psychological Science 24~(3) (2015) 163--169.

\bibitem{orrell2018quantumbooknewscience}
D.~Orrell, Quantum economics: The new science of money, Icon Books, 2018.

\bibitem{Orrell:MoneyMagic}
D.~Orrell, Money, Magic, and How to Dismantle a Financial Bomb: Quantum Economics for the Real World, Icon Books Ltd, 39–41 North Road, London N7 9DP, 2022.

\bibitem{pothos2022quantum}
E.~M. Pothos, J.~R. Busemeyer, Quantum cognition, Annual review of psychology 73 (2022) 749--778.

\bibitem{hampton1988disjunction}
J.~A. Hampton, Disjunction of natural concepts, Memory \& Cognition 16~(6) (1988) 579--591.

\bibitem{shafffi1990typicality}
E.~B. Shafffi, E.~E. Smith, D.~N. Osherson, Typicality and reasoning fallacies, Memory \& Cognition 18~(3) (1990) 229--239.

\bibitem{tentori2004conjunction}
K.~Tentori, N.~Bonini, D.~Osherson, The conjunction fallacy: A misunderstanding about conjunction?, Cognitive Science 28~(3) (2004) 467--477.

\bibitem{aerts2012quantum}
D.~Aerts, S.~Sozzo, Quantum interference in cognition: Structural aspects of the brain, arXiv preprint arXiv:1204.4914 (2012).

\bibitem{orrell2020supplyanddemand}
D.~Orrell, A quantum model of supply and demand, Physica A: Statistical Mechanics and its Applications 539 (2020) 122928.

\bibitem{orrell2022quantum}
D.~Orrell, Quantum impact and the supply-demand curve, Available at SSRN (2022).

\bibitem{ahn2024business}
K.~Ahn, L.~Cong, H.~Jang, D.~S. Kim, Business cycle and herding behavior in stock returns: theory and evidence, Financial Innovation 10~(1) (2024) 6.

\bibitem{meng2016quantum}
X.~Meng, J.-W. Zhang, H.~Guo, Quantum {Brownian} motion model for the stock market, Physica A: Statistical Mechanics and its Applications 452 (2016) 281--288.

\bibitem{orus2019quantum}
R.~Or{\'u}s, S.~Mugel, E.~Lizaso, Quantum computing for finance: overview and prospects, Reviews in Physics 4 (2019) 100028.

\bibitem{egger2020quantum}
D.~J. Egger, C.~Gambella, J.~Marecek, S.~McFaddin, M.~Mevissen, R.~Raymond, A.~Simonetto, S.~Woerner, E.~Yndurain, Quantum computing for finance: State-of-the-art and future prospects, IEEE Transactions on Quantum Engineering 1 (2020) 1--24.

\bibitem{herman2022survey}
D.~Herman, C.~Googin, X.~Liu, A.~Galda, I.~Safro, Y.~Sun, M.~Pistoia, Y.~Alexeev, A survey of quantum computing for finance, arXiv preprint arXiv:2201.02773 (2022).

\bibitem{widdows2023quantum}
D.~Widdows, J.~Rani, E.~M. Pothos, Quantum circuit components for cognitive decision-making, Entropy 25~(4) (2023) 548.

\bibitem{widdows2024quantum}
D.~Widdows, A.~Bhattacharyya, Quantum financial modeling on noisy intermediate-scale quantum hardware: Random walks using approximate quantum counting, Quantum Economics and Finance 1~(1) (2024) 5--20.

\bibitem{puengtambol2021implementation}
W.~Puengtambol, P.~Prechaprapranwong, U.~Taetragool, Implementation of quantum random walk on a real quantum computer, Journal of Physics: Conference Series 1719 (2021) 012103.

\bibitem{venegas2012quantum}
S.~E. Venegas-Andraca, Quantum walks: a comprehensive review, Quantum Information Processing 11~(5) (2012) 1015--1106.

\bibitem{BlackScholes:1973}
F.~Black, M.~Scholes, The {Pricing} of {Options} and {Corporate} {Liabilities}, Journal of Political Economy 81~(3) (1973) 637--654.

\bibitem{ziemann2021physics}
V.~Ziemann, \href{https://books.google.be/books?id=NvoVEAAAQBAJ}{Physics and Finance}, Undergraduate Lecture Notes in Physics, Springer International Publishing, 2021.
\newline\urlprefix\url{https://books.google.be/books?id=NvoVEAAAQBAJ}

\bibitem{mandelbrot1963variationcertain}
B.~Mandelbrot, The {Variation} of {Certain} {Speculative} {Prices}, The Journal of Physics 36~(4) (1963) 394--419.

\bibitem{gabaix2009power}
X.~Gabaix, Power {Laws} in {Economics} and {Finance}, Annual Review of Economics 1~(1) (2009) 255--294.

\bibitem{gopikrishnan1998inverse}
P.~Gopikrishnan, M.~Meyer, L.~N. Amaral, H.~E. Stanley, Inverse cubic law for the distribution of stock price variations, The European Physical Journal B-Condensed Matter and Complex Systems 3~(2) (1998) 139--140.

\bibitem{gopikrishnan1999scaling}
P.~Gopikrishnan, V.~Plerou, L.~A.~N. Amaral, M.~Meyer, H.~E. Stanley, Scaling of the distribution of fluctuations of financial market indices, Physical Review E 60~(5) (1999) 5305.

\bibitem{Bachelier:1900}
L.~Bachelier, Theorie de la spéculation, Annales Scientifiques de l’École Normale Supérieure 3~(17) (1900) 21--86.

\bibitem{orrell2020quantummathsbook}
D.~Orrell, Quantum Economics and Finance: An Applied Mathematics Introduction, Panda Ohana Publishing, 2020.

\bibitem{mandelbrot1997variation}
B.~Mandelbrot, The variation of the prices of cotton, wheat, and railroad stocks, and of some financial rates, Fractals and Scaling in Finance: Discontinuity, Concentration, Risk. Selecta Volume E (1997) 419--443.

\bibitem{plerou2000econophysics}
V.~Plerou, P.~Gopikrishnan, B.~Rosenow, L.~A. Amaral, H.~E. Stanley, Econophysics: financial time series from a statistical physics point of view, Physica A: Statistical Mechanics and its Applications 279~(1-4) (2000) 443--456.

\bibitem{mao2007stochastic}
X.~Mao, Stochastic differential equations and applications, Elsevier, 2007.

\bibitem{yarahmadi20222d}
H.~Yarahmadi, A.~A. Saberi, A {2D} {L\'evy}-flight model for the complex dynamics of real-life financial markets, Chaos: An Interdisciplinary Journal of Nonlinear Science 32~(3) (2022).

\bibitem{ding1993long}
Z.~Ding, C.~W. Granger, R.~F. Engle, A long memory property of stock market returns and a new model, Journal of empirical finance 1~(1) (1993) 83--106.

\bibitem{kou2002jump}
S.~G. Kou, A jump-diffusion model for option pricing, Management science 48~(8) (2002) 1086--1101.

\bibitem{malkiel2003efficient}
B.~G. Malkiel, The efficient market hypothesis and its critics, Journal of economic perspectives 17~(1) (2003) 59--82.

\bibitem{Holt:2013}
J.~Holt, A {Random} {Walk} with {Louis} {Bachelier}, \url{https://www.nybooks.com/articles/2013/10/24/random-walk-louis-bachelier/}, accessed: 2023-12-08 (2013).

\bibitem{Scanlon:2019}
K.~Scanlon, Brownian {Motion}, {Random} {Walks}, and the {Hot} {Hands} {Fallacy}, \url{https://medium.com/analytics-vidhya/brownian-motion-random-walks-and-the-hot-hands-fallacy-e24aba54804f}, accessed: 2023-12-08 (2019).

\bibitem{lekovic2018evidence}
M.~Lekovi{\'c}, Evidence for and against the validity of efficient market hypothesis, Economic themes 56~(3) (2018) 369--387.

\bibitem{fama1963mandelbrot}
E.~F. Fama, Mandelbrot and the stable {Paretian} hypothesis, The journal of business 36~(4) (1963) 420--429.

\bibitem{mittnik1999maximum}
S.~Mittnik, T.~Doganoglu, D.~Chenyao, et~al., Maximum likelihood estimation of stable {Paretian} models, Mathematical and Computer modelling 29~(10-12) (1999) 275--293.

\bibitem{mandelbrot1966persistence}
B.~Mandelbrot, Is there persistence in stock price movements?, in: Seminar on the Analysis of Security Prices. Graduate Scholl of Business: University of Chicago, 1966, pp. 1--3.

\bibitem{madan1990variance}
D.~B. Madan, E.~Seneta, The variance gamma ({VG}) model for share market returns, Journal of business (1990) 511--524.

\bibitem{kiyono2006power}
K.~Kiyono, Z.~R. Struzik, Y.~Yamamoto, Power law and its transition in the slow convergence to a {Gaussian} in the {S\&P500} index, in: Practical Fruits of Econophysics: Proceedings of the Third Nikkei Econophysics Symposium, Springer, 2006, pp. 67--71.

\bibitem{tuncay2007power}
{\c{C}}.~Tuncay, D.~Stauffer, Power laws and {Gaussians} for stock market fluctuations, Physica A: Statistical Mechanics and its Applications 374~(1) (2007) 325--330.

\bibitem{kiyono2006criticality}
K.~Kiyono, Z.~R. Struzik, Y.~Yamamoto, Criticality and phase transition in stock-price fluctuations, Physical review letters 96~(6) (2006) 068701.

\bibitem{matia2004scale}
K.~Matia, M.~Pal, H.~Salunkay, H.~E. Stanley, Scale-dependent price fluctuations for the {Indian} stock market, Europhysics Letters 66~(6) (2004) 909.

\bibitem{zhang2007power}
J.~Zhang, Y.~Zhang, H.~Kleinert, Power tails of index distributions in {Chinese} stock market, Physica A: Statistical Mechanics and its Applications 377~(1) (2007) 166--172.

\bibitem{newman2005power}
M.~E. Newman, Power laws, {Pareto} distributions and {Zipf's} law, Contemporary physics 46~(5) (2005) 323--351.

\bibitem{gabaix2003theory}
X.~Gabaix, P.~Gopikrishnan, V.~Plerou, H.~E. Stanley, A theory of power-law distributions in financial market fluctuations, Nature 423~(6937) (2003) 267--270.

\bibitem{farmer2004origin}
J.~D. Farmer, F.~Lillo, On the origin of power-law tails in price fluctuations, Quantitative Finance 4~(1) (2004) C7.

\bibitem{mitzenmacher2004brief}
M.~Mitzenmacher, A brief history of generative models for power law and lognormal distributions, Internet mathematics 1~(2) (2004) 226--251.

\bibitem{markovic2014power}
D.~Markovi{\'c}, C.~Gros, Power laws and self-organized criticality in theory and nature, Physics Reports 536~(2) (2014) 41--74.

\bibitem{yuan2018cev}
W.~Yuan, S.~Lai, The {CEV} model and its application to financial markets with volatility uncertainty, Journal of Computational and Applied Mathematics 344 (2018) 25--36.

\bibitem{rogers1997arbitrage}
L.~C.~G. Rogers, Arbitrage with fractional {Brownian} motion, Mathematical finance 7~(1) (1997) 95--105.

\bibitem{rostek2013note}
S.~Rostek, R.~Sch{\"o}bel, A note on the use of fractional {Brownian} motion for financial modeling, Economic Modelling 30 (2013) 30--35.

\bibitem{merton1976option}
R.~C. Merton, Option pricing when underlying stock returns are discontinuous, Journal of financial economics 3~(1-2) (1976) 125--144.

\bibitem{hanson2002jump}
F.~Hanson, J.~Westman, Jump-{Diffusion} {Stock} {Return} {Models} in {Finance}: {Stochastic} {Process} {Density} with {Uniform}-{Jump} {Amplitude}, Proceedings of the 15th International Symposium on Mathematical Theory of Networks and Systems (01 2002).

\bibitem{cont1997scaling}
R.~Cont, M.~Potters, J.-P. Bouchaud, Scaling in stock market data: stable laws and beyond, in: Scale Invariance and Beyond: Les Houches Workshop, March 10--14, 1997, Springer, 1997, pp. 75--85.

\bibitem{eberlein1995hyperbolic}
E.~Eberlein, U.~Keller, Hyperbolic distributions in finance, Bernoulli (1995) 281--299.

\bibitem{barndorff1997normal}
O.~E. Barndorff-Nielsen, Normal inverse {Gaussian} distributions and stochastic volatility modelling, Scandinavian Journal of statistics 24~(1) (1997) 1--13.

\bibitem{blattberg2010comparison}
R.~C. Blattberg, N.~J. Gonedes, A comparison of the stable and student distributions as statistical models for stock prices, in: Perspectives on promotion and database marketing: The collected works of Robert C Blattberg, World Scientific, 2010, pp. 25--61.

\bibitem{laherrere1998stretched}
J.~Laherrere, D.~Sornette, Stretched exponential distributions in nature and economy: “fat tails” with characteristic scales, The European Physical Journal B-Condensed Matter and Complex Systems 2 (1998) 525--539.

\bibitem{malevergne2005empirical}
Y.~Malevergne, V.~Pisarenko, D.~Sornette, Empirical distributions of stock returns: between the stretched exponential and the power law?, Quantitative Finance 5~(4) (2005) 379--401.

\bibitem{orrell2021quantumwalkArticle}
D.~Orrell, A quantum walk model of financial options, Wilmott 2021~(112) (2021) 62--69.

\bibitem{kempe2003quantum}
J.~Kempe, Quantum random walks: an introductory overview, Contemporary Physics 44~(4) (2003) 307--327.

\bibitem{chandrashekar2008optimizing}
C.~M. Chandrashekar, R.~Srikanth, R.~Laflamme, Optimizing the discrete time quantum walk using a {SU} (2) coin, Physical Review A 77~(3) (2008) 032326.

\bibitem{reitzner2012quantum}
D.~Reitzner, D.~Nagaj, V.~Buzek, Quantum walks, arXiv preprint arXiv:1207.7283 (2012).

\bibitem{aharonov1993quantum}
Y.~Aharonov, L.~Davidovich, N.~Zagury, Quantum random walks, Physical Review A 48~(2) (1993) 1687.

\bibitem{jayakody2023revisiting}
M.~N. Jayakody, C.~Meena, P.~Pradhan, Revisiting one-dimensional discrete-time quantum walks with general coin, Physics Open 17 (2023) 100189.

\bibitem{romanelli2007measurements}
A.~Romanelli, Measurements in the {L\'evy} quantum walk, Physical Review A 76~(5) (2007) 054306.

\bibitem{ishak2021entropy}
N.~I. Ishak, S.~V. Muniandy, W.~Y. Chong, Entropy analysis of the discrete-time quantum walk under bit-flip noise channel, Physica A: Statistical Mechanics and its Applications 584 (2021) 126371.

\bibitem{farmer2000physicists}
J.~D. Farmer, Physicists attempt to scale the ivory towers of finance, International Journal of Theoretical and Applied Finance 3~(03) (2000) 311--333.

\bibitem{cont2007volatility}
R.~Cont, Volatility clustering in financial markets: empirical facts and agent-based models, in: Long memory in economics, Springer, 2007, pp. 289--309.

\bibitem{romanelli2004quantum}
A.~Romanelli, A.~S. Schifino, R.~Siri, G.~Abal, A.~Auyuanet, R.~Donangelo, Quantum random walk on the line as a {Markovian} process, Physica A: Statistical Mechanics and its Applications 338~(3-4) (2004) 395--405.

\bibitem{sornette2018can}
D.~Sornette, P.~Cauwels, G.~Smilyanov, Can we use volatility to diagnose financial bubbles? {Lessons} from 40 historical bubbles, Quantitative Finance and Economics 2~(1) (2018) 486--594.

\bibitem{pires2020quantum}
M.~A. Pires, S.~D. Queir{\'o}s, Quantum walks with sequential aperiodic jumps, Physical Review E 102~(1) (2020) 012104.

\bibitem{ebrahimi1999ordering}
N.~Ebrahimi, E.~Maasoumi, E.~S. Soofi, Ordering univariate distributions by entropy and variance, Journal of Econometrics 90~(2) (1999) 317--336.

\bibitem{gamble2009demystifying}
J.~K. Gamble, J.~F. Lindner, Demystifying decoherence and the master equation of quantum {Brownian} motion, American Journal of Physics 77~(3) (2009) 244--252.

\bibitem{schlosshauer2019quantum}
M.~Schlosshauer, Quantum decoherence, Physics Reports 831 (2019) 1--57.

\bibitem{romanelli2005decoherence}
A.~Romanelli, R.~Siri, G.~Abal, A.~Auyuanet, R.~Donangelo, Decoherence in the quantum walk on the line, Physica A: Statistical Mechanics and its Applications 347 (2005) 137--152.

\bibitem{romanelli2011coinflipping}
A.~Romanelli, G.~Hern{\'a}ndez, Quantum walks: Decoherence and coin-flipping games, Physica A: Statistical Mechanics and its Applications 390~(6) (2011) 1209--1220.

\bibitem{mackay2002quantum}
T.~D. Mackay, S.~D. Bartlett, L.~T. Stephenson, B.~C. Sanders, Quantum walks in higher dimensions, Journal of Physics A: Mathematical and General 35~(12) (2002) 2745.

\bibitem{brun2003quantumDecoherentCoins}
T.~A. Brun, H.~A. Carteret, A.~Ambainis, Quantum random walks with decoherent coins, Physical Review A 67~(3) (2003) 032304.

\bibitem{brun2003quantumManyCoins}
T.~A. Brun, H.~A. Carteret, A.~Ambainis, Quantum walks driven by many coins, Physical Review A 67~(5) (2003) 052317.

\bibitem{jayakody2018transfiguration}
M.~N. Jayakody, A.~Nanayakkara, Transfiguration of {Quantum} {Walks} on a line, arXiv preprint arXiv:1809.00505 (2018).

\bibitem{sheraz2015entropy}
M.~Sheraz, S.~Dedu, V.~Preda, Entropy measures for assessing volatile markets, Procedia Economics and Finance 22 (2015) 655--662.

\bibitem{drzazga2023entropy}
E.~A. Drzazga-Szcz\c{e}{\'s}niak, P.~Szczepanik, A.~Z. Kaczmarek, D.~Szcz\c{e}{\'s}niak, Entropy of financial time series due to the shock of war, Entropy 25~(5) (2023) 823.

\end{thebibliography}

\end{document}